\documentclass[pra,superscriptaddress,twocolumn,showpacs]{revtex4-1}

\usepackage[commandnameprefix=always]{changes}

\usepackage{latexsym}
\usepackage{graphicx}
\usepackage[colorlinks=true, citecolor=blue, urlcolor=blue]{hyperref}
\usepackage{float}
\usepackage{textcomp}
\usepackage{mathpazo}
\usepackage{comment}
\usepackage{lipsum}

\usepackage{exscale}
\usepackage{bbm}
\usepackage{latexsym}
\usepackage[T1]{fontenc}
\usepackage{enumerate}
\usepackage{bbold}
\usepackage{color}
\usepackage[colorlinks=true,citecolor=blue,urlcolor=blue]{hyperref}
\usepackage{mathtools}
\usepackage{amsfonts,amsmath,amssymb,amsthm}

\usepackage[normalem]{ulem}

\usepackage{xr}
\makeatletter
\newcommand*{\addFileDependency}[1]{
  \typeout{(#1)}
  \@addtofilelist{#1}
  \IfFileExists{#1}{}{\typeout{No file #1.}}
}
\makeatother

\newcommand*{\myexternaldocument}[1]{
    \externaldocument{#1}
    \addFileDependency{#1.tex}
    \addFileDependency{#1.aux}
}

\myexternaldocument{appendix}

\usepackage{bbm}

\usepackage[colorlinks=true,citecolor=blue,urlcolor=blue]{hyperref}
\usepackage[draft]{fixme}
\usepackage{amsmath,bbm}
\usepackage{graphicx}
\usepackage{amsfonts}
\usepackage{amssymb}
\usepackage{amsmath, amssymb, amsthm,verbatim,graphicx,bbm}
\usepackage{mathrsfs}
\usepackage{color,xcolor,longtable}
\usepackage{changes}

\newcommand{\beq}[0]{\begin{equation}}
\newcommand{\eeq}[0]{\end{equation}}

\newcommand{\one}{\leavevmode\hbox{\small1\normalsize\kern-.33em1}}

\def\be{\begin{equation}}
\def\ee{\end{equation}}
\def\ben{\begin{eqnarray}}
\def\een{\end{eqnarray}}
\def\eea{\end{array}}
\def\bea{\begin{array}}

\newcommand{\Tr}[1]{\mathrm{Tr}#1}
\newcommand{\bei}{\begin{itemize}}
\newcommand{\eei}{\end{itemize}}
\newcommand{\ket}[1]{|#1\rangle}
\newcommand{\bra}[1]{\langle#1|}

\newcommand{\proj}[1]{\ket{#1}\!\bra{#1}}

\newcommand{\I}{\mathbbm{1}}

\renewcommand{\emph}[1]{\textbf{#1}}

\newcommand{\eqnref}[1]{(\ref{#1})}

\makeatletter
\newtheorem*{rep@theorem}{\rep@title}
\newcommand{\newreptheorem}[2]{%
\newenvironment{rep#1}[1]{%
 \def\rep@title{#2 \ref{##1}}%
 \begin{rep@theorem}}%
 {\end{rep@theorem}}}
\makeatother

\theoremstyle{plain}
\newtheorem{thm}{Theorem}
\newtheorem*{thm*}{Theorem}
\newreptheorem{thm}{Theorem}

\newtheorem{fakt}{Fact}

\newtheorem{obs}{Observation}[thm]

\newtheorem{cor}{Corollary}

\theoremstyle{definition}

\theoremstyle{remark}

\usepackage[T1]{fontenc}

\begin{document}


\title{Self-testing of any pure entangled state with minimal number of measurements
and optimal randomness certification in one-sided device-independent scenario}



\author{Shubhayan Sarkar}
\affiliation{Center for Theoretical Physics, Polish Academy of Sciences, Aleja Lotnik\'{o}w 32/46, 02-668 Warsaw, Poland}
\author{Jakub J. Borka\l{}a}
\affiliation{Center for Theoretical Physics, Polish Academy of Sciences, Aleja Lotnik\'{o}w 32/46, 02-668 Warsaw, Poland}
\author{Chellasamy Jebarathinam}
\affiliation{Center for Theoretical Physics, Polish Academy of Sciences, Aleja Lotnik\'{o}w 32/46, 02-668 Warsaw, Poland}

\author{Owidiusz Makuta}
\affiliation{Center for Theoretical Physics, Polish Academy of Sciences, Aleja Lotnik\'{o}w 32/46, 02-668 Warsaw, Poland}
\author{Debashis Saha}
\affiliation{Center for Theoretical Physics, Polish Academy of Sciences, Aleja Lotnik\'{o}w 32/46, 02-668 Warsaw, Poland}

\affiliation{S. N. Bose National Centre for Basic Sciences,
Block JD, Sector III, Salt Lake, Kolkata 700098, India }

\author{Remigiusz Augusiak}
\affiliation{Center for Theoretical Physics, Polish Academy of Sciences, Aleja Lotnik\'{o}w 32/46, 02-668 Warsaw, Poland}

\begin{abstract}	

Certification of quantum systems and their properties has become a field of intensive studies.
%
%
Here, taking advantage of the one-sided device-independent scenario (known also as quantum steering scenario), we propose a self-testing scheme for all bipartite entangled states using a single family of steering inequalities with the minimal number of two measurements per party. Building on this scheme we then show how to certify all rank-one extremal measurements, including non-projective $d^2$-outcome measurements, which in turn can be used for certification of the maximal amount of randomness,
 that is, $2\log_2d$ bits. 
Finally, in a particular case of $d=3$, we extend our self-testing results to the fully device-independent setting.

\end{abstract}


\maketitle


\textit{Introduction.---}Randomness plays an essential role in numerous fields such as simulations, cryptography and sampling \cite{PAB+20}. Interestingly, measurements on quantum systems can be used to generate true randomness. One of the most secure scenarios for random number generation is based on quantum non-locality \cite{BCP+14,CS16}. Recently, quantum non-locality, which can be observed via the violation of Bell inequalities \cite{bel64}, has been identified as a resource for certifying randomness in a device-independent (DI) way, that is, independent of the detailed physical description of the experimental set-up \cite{AM16}. However, most of the known results in the DI setting concern composite quantum systems of relatively low dimensions \cite{random0,random1,Tavakoli2021}. In fact, there exist only a few protocols for randomness certification in systems of arbitrary dimension; see, e.g., Ref. \cite{sarkar} where such a scheme for certification of $\log_2d$ random bits based on the maximally entangled states of local dimension $d$ and projective measurements was proposed.
%
%
It remains a highly non-trivial problem to find methods of optimal randomness certification using any bipartite entangled state and non-projective measurements. 

From an experimental point of view, implementation of Bell nonlocality is challenging due to the requirement of low levels of noise and high detector efficiencies. On the other hand, the observation of asymmetric form of quantum nonlocality, known as quantum steering \cite{WJD07}, can be revealed through violation of steering inequalities \cite{CJWR09}, and is more robust to noise and imperfections of the detectors \cite{PCS+15}.  Motivated by this, quantum steering has been intensively studied as a resource for various protocols in a one-sided device-independent ($1$SDI) scenario, in which a part of the experimental set-up is assumed to be characterized and trusted \cite{BCW+12,LTB+14,Supic}. Recently, the possibilities for certification of quantum states and measurements in 1SDI scenario were investigated in Refs. \cite{Supic, Alex, Goswami, SBK20, Chen, sarkar2}.
Moreover, the growing interest in practical quantum randomness certification provides additional motivation to look for scenarios that are easier to implement and have the highest possible efficiency, that is, those that employ minimal resources in terms of states and measurements and exhibit high noise resistance \cite{FVM+19,KST+19,GCH+19,DSU+21}. In fact, randomness generation within the 1SDI scenario has recently been studied \cite{PCS+15,Dani2018}; in particular Ref. \cite{Dani2018} provides a scheme for the generation of maximal amount of $\log_2d$ with the aid
of $d$-outcome projective measurements. However, it remains an open and highly nontrivial problem whether one can generate the maximal amount of randomness achievable in quantum systems of dimension $d$, i.e., $2\log_2d$ bits.   

This work aims to solve the above problem and takes advantage of the one-sided device-independent setting to propose a scheme for optimal randomness certification in quantum systems of arbitrary local dimension. 
The first step we make on the way to achieve this result is to construct a class of steering inequalities that are maximally violated by any pure entangled two qudit state and just two measurements on each side, which is the minimum number needed to observe quantum steering. Importantly, we then show that maximal violation of these inequalities can be used for 1SDI certification of any two-qudit pure entangled state and mutually unbiased measurements on Bob's side. Additionally, we show that any rank-one extremal measurement can be certified using our scheme. Based on the above results, we present a certification method of $2\log_2d$ bits of randomness from maximally entangled state of any local dimension $d$ and also from some partially entangled states of local dimension $3,4,5$ and $6$. 
In the particular case of $d=3$, we finally extend our self-testing results to the device-independent setting. 
%

\textit{Preliminaries.---}We first provide some background information necessary for further
considerations.

\textit{The scenario.} We consider the $1$SDI scenario consisting of
two parties Alice and Bob, located in separated laboratories that share some quantum state $\rho_{AB}$. 
%
%
Both parties have access to measuring devices which they use to perform measurements on their shares of $\rho_{AB}$. Contrary to the device-independent scenario, here we assume that one of the measuring devices is trusted (say, the one held by Alice) and performs known measurements. Yet, unlike in the standard steering scenario, we do not need to assume that her device performs full quantum tomography. On the other hand, Bob's device remains untrusted and can be treated as a 'black box'. The measurement choices of Alice and Bob are denoted  $x$ and $y$, whereas the obtained outcomes $a$ and $b$, respectively.


The correlations obtained in this scenario are captured by a set of probability distributions $\vec{p}=\{p(a,b|x,y)\}$, where $p(a,b|x,y)$ is the probability of observing outcomes $a$
and $b$ by Alice and Bob after performing measurements $x$ and $y$, respectively; it expresses as
\begin{equation}
    p(a,b|x,y)=\Tr\left[\rho_{AB}(M_{a|x}\otimes N_{b|y})\right],
\end{equation}
where $\rho_{AB}$ is the state shared by Alice and Bob, and $M_x=\{M_{a|x}\}$ and
$N_y=\{N_{b|y}\}$ are Alice's and Bob's measurements; recall that the measurement operators $M_{a|x}$ and $N_{b|y}$ are positive semi-definite and 
sum up to the identity on the corresponding Hilbert space for each $x$ and $y$. 
In what follows, we assume that Alice performs projective measurements; then $M_{a|x}$ are pairwise orthogonal projections. On the other hand, we make no assumptions about Bob's measurements or the shared state; in fact, we consider a fully general situation of $\rho_{AB}$ being mixed and Bob's measurement being POVM's.
%

In $1$SDI scenario, quantum steering is demonstrated if the probability distribution $\{p(a,b|x,y)\}$ cannot be described by a local hidden state (LHS) model \cite{WJD07}, that is,
\begin{equation}
    p(a,b|x,y)=\sum_\lambda p(\lambda) \Tr [M_{a|x}\, \rho^\lambda_A] p(b|y,\lambda),
\end{equation}
where $\lambda$ is a hidden variable distributed according to the probability distribution $p(\lambda)$ and determines both the state $\rho^\lambda_A$ sent to Alice and Bob's response $p(b|y,\lambda)$. Any set of probability distribution $\{p(a,b|x,y)\}$ which does not admit the above form can be detected through the violation of a linear steering inequality,
$\mathcal{B}[p(a,b|x,y)] \le \beta_{L}$, where $\mathcal{B}$ is a linear combination of the probabilities $p(a,b|x,y)$, 
%
%
where $\beta_{L}$ is the classical or LHS bound \cite{CJWR09}. 

It is beneficial to express 
the observed correlations in terms of the 
generalized expectation values defined as
%
%
\begin{equation}
    \langle A_{k|x} B_{l|y} \rangle = \sum^{d-1}_{a,b=0} \omega^{ak+bl} p(a,b|x,y),
\end{equation}
where $\omega=\exp(2\pi\mathbbm{i}/d)$ and $k,l=0,\ldots,d-1$. If $p(a,b|x,y)$ are quantum, 
these can be expressed as $\langle A_{k|x} B_{l|y} \rangle=\Tr[(A_x^k\otimes B_{l|y})\rho_{AB}]$, where $A_x^k$ are $k$th powers of a unitary observable $A_x$ ($A_x^d=\mathbbm{1}_d$), whereas $B_{l|y}$ are Fourier transforms of the measurement operators $N_{b|y}$: $B_{l|y}=\sum_{b}\omega^{lb}N_{b|y}$. In fact, the operators $B_{l|y}$ provide an alternative, but unique representation of the corresponding measurement $N_y$. We can thus refer to the set $B_y=\{B_{l|y}\}_{l=0}^{d-1}$ as Bob's measurement too. If $N_y$ is projective, then $B_{l|y}$ are $l$th powers of a unitary observable, which, slightly abusing the notation, we also denote $B_y$ $(B_y^d=\mathbbm{1})$, that is, $B_{l|y}=B_y^l$ (see Appendix \ref{App0}).



\textit{Randomness certification.} In order to certify true randomness in Bob's outcomes, we consider an adversarial scenario, where an eavesdropper Eve has access to Bob's laboratory. We therefore assume the state $\rho_{AB}$ to be a reduced state of a (in general entangled) state $\rho_{ABE}$ shared by Alice, Bob and Eve, i.e. $\rho_{AB} = \Tr_E [ \rho_ {ABE} ]$. We can safely assume it to be pure. Now, Eve's aim is to correctly guess the outcome of one of Bob's measurements, say $N_y$. To do so, on her share of $\rho_{ABE}$ she performs a measurement $Z=\{E_e\}$ whose outcome $e$ is the best guess on Bob's outcome. We quantify the amount of random bits obtained from Bob's measurement with $H_{\min}=-\log_2 G(y,\vec{p})$, where $G(y,\vec{p})$ is the local guessing probability defined as
\begin{equation}\label{LGpr}
    G(y,\vec{p})=\sup_{S\in S_{\vec{p}}}\sum_{b}\Tr\left[\rho_{ABE}\left( \I \otimes N_{b|y} \otimes E_b \right)\right],
\end{equation}
where $S_{\vec{p}}$ is the set of all Eve's strategies (consisting of the shared state $\rho_{ABE}$ and the measurement $Z$) that reproduce the correlations $\vec{p}$ observed by Alice and Bob.
%
%
As was demonstrated in \cite{PCS+15}, quantum realisations that violate steering inequalities can be used to certify genuine randomness on the untrusted side. One of our aims here is to show that the maximal amount of $2\log_2d$ random bits can be certified from pure entangled bipartite states of Schmidt rank $d$ in the 1SDI scenario.

\textit{Results.---}We first introduce a general class of steering inequalities that are maximally violated by any two-qudit pure entangled state and two measurements on both sides. Consider an entangled state from $\mathbbm{C}^d\otimes \mathbbm{C}^d$
in its Schmidt decomposition
\begin{eqnarray}\label{theState}
\ket{\psi(\boldsymbol{\alpha})}_{AB}=\sum_{i=0}^{d-1}\alpha_{i}\ket{i}_A\ket{i}_B,
\end{eqnarray}
where $\boldsymbol{\alpha}$ is a vector composed of the Schmidt coefficients $\alpha_i\geq 0$
satisfying $\alpha_0^2+\ldots+\alpha_{d-1}^2=1$.
Clearly, without any loss of generality,
we can fix the local bases to be the computational basis of $\mathbbm{C}^d$. 
We can also assume that all $\alpha_i>0$, or, equivalently, 
that the Schmidt rank of $\ket{\psi(\boldsymbol{\alpha})}$ is maximal;
otherwise $\ket{\psi(\boldsymbol{\alpha})}$ is not a two-qudit state. 

Now, our steering inequality maximally violated by  
(\ref{theState}) for a given $\boldsymbol{\alpha}$
can be stated in the observable picture as
\begin{equation}\label{Steineq1}
\mathcal{B}_d(\boldsymbol{\alpha})\!\!=\!\!\sum_{k=1}^{d-1}\left\langle A_0^{k}\otimes B_{k|0} \!+\! 
\gamma(\boldsymbol{\alpha}) A_1^{k}\otimes B_{k|1}\!+\!\delta_k(\boldsymbol{\alpha})A_0^{k}\right\rangle
\leq \beta_L,
%
\end{equation}
where $\gamma(\boldsymbol{\alpha})$ and $\delta_k(\boldsymbol{\alpha})$ are $d$ coefficients given by 
\begin{eqnarray}
%
\gamma(\boldsymbol{\alpha})=\frac{d}{\sum_{\substack{i,j=0\\i\ne j}}^{d-1}\frac{\alpha_i}{\alpha_j}},\quad \delta_k(\boldsymbol{\alpha})=-\frac{\gamma}{d}\sum_{\substack{i,j=0\\i\ne j}}^{d-1}\frac{\alpha_i}{\alpha_j}\omega^{k(d-j)}.
\end{eqnarray}
Whereas Bob performs two arbitrary measurements $B_0$ and $B_1$, Alice's observables 
are known and fixed to be simply $A_0=Z_d$ and $A_1=X_d$, where
\begin{eqnarray}\label{Z,X}
Z_d=\sum_{i=0}^{d-1}\omega^i\ket{i}\!\bra{i},\qquad X_d=\sum_{i=0}^{d-1}\ket{i+1}\!\bra{i}
\end{eqnarray}  
are generalizations of Pauli matrices to $d$-dimensional Hilbert spaces. It should be mentioned 
that an alternative construction of steering inequalities maximally violated by
two-qudit pure entangled states is given in Ref. \cite{Dani2018}, 
however, except for the particular case of the maximally entangled state, those inequalities require $d+1$ measurements on the trusted side to observe the maximal quantum violation.
Our inequalities require only two measurements on the trusted side for any state (\ref{theState}).

While it is in general challenging to determine the classical bound $\beta_L(\boldsymbol{\alpha})$
for any $\boldsymbol{\alpha}$, one can 
derive (see Appendix \ref{App A} for details) the following upper bound on it:
\begin{eqnarray}\label{classbound}
   %
   \mathcal{B}_L(\boldsymbol{\alpha})&\leq& \max_{\ket{\phi}}\left\{d\max_{a}\{|\eta_a|^2\}+\gamma(\boldsymbol{\alpha})\left(\sum_{a=0}^{d-1}|\eta_a|\right)^2\right.\nonumber\\
    &&\hspace{1cm}\left.-\gamma(\boldsymbol{\alpha})\sum_{i,a=0}^{d-1}\alpha_i\frac{|\eta_a|^2}{\alpha_a}\right\},
\end{eqnarray}
where the maximization is over $d$ numbers $|\eta_i|$ obeying
$|\eta_0|^2+\ldots+|\eta_{d-1}|^2=1$, which come from the decomposition of some quantum state
$\ket{\phi}=\sum_{i}\eta_i\ket{i}\in\mathbbm{C}^d$.

On the other hand, the maximal quantum value of $\mathcal{B}_d(\boldsymbol{\alpha})$ 
can be analytically found for any choice of $\alpha_i>0$ and reads $\beta_Q(\boldsymbol{\alpha})=d$ (see Appendix \ref{app:SOS} for the proof). 
Crucially, this value is achieved by the state (\ref{theState}) 
and Bob's measurements being simply $B_0=Z_d^*$ and $B_1=X_d$. 
For a remark, it is interesting to notice that for this choice of observables 
the steering operator corresponding to (\ref{Steineq1}) becomes simply a combination of two stabilizing operators that stabilize the state (\ref{theState}). Moreover, 
$\beta_Q(\boldsymbol{\alpha})$ is strictly larger than $\beta_{L}(\boldsymbol{\alpha})$ for any 
$\boldsymbol{\alpha}$ such that each $\alpha_i>0$.
\begin{thm}
For any set of positive and normalized coefficient $\alpha_i$,
the maximal quantum value of $\mathcal{B}_d(\boldsymbol{\alpha})$ is 
$\beta_Q(\boldsymbol{\alpha})=d$ and obeys $\beta_Q(\boldsymbol{\alpha})>\beta_L(\boldsymbol{\alpha})$.
\end{thm}

We thus have a general class of nontrivial steering inequalities maximally violated by any pure entangled state (\ref{theState}) and two measurements on Bob's side, which is, in fact, the minimal number necessary to observe steerability. Moreover, as we show in the following theorem (see Appendix \ref{App C} for a proof), their maximal violation can be used for 1SDI self-testing of any entangled state (\ref{theState}).
\begin{thm}\label{Theo1} 
Assume that the steering inequality \eqref{Steineq1} with $A_0=Z_d$ and $A_1=X_d$ is maximally violated by $\ket{\psi}_{ABE}\in(\mathbbm{C}^d)_A\otimes\mathcal{H}_B\otimes\mathcal{H}_E$ and Bob's measurements $B_i$ $(i=1,2)$. Then, the following statements hold true for any $d$: (i) $\mathcal{H}_B=(\mathbbm{C}^d)_{B'}\otimes\mathcal{H}_{B''}$, where $\mathcal{H}_{B''}$ is some finite-dimensional Hilbert space, (ii) Bob's measurements are projective and there is a local unitary transformation $U_B:\mathcal{H}_B\rightarrow\mathcal{H}_B$, such that
\begin{eqnarray}
\forall i, \quad U_B\,B_i\,U_B^{\dagger}=A_i^{*}\otimes\I_{B''},
\end{eqnarray}
where $\mathbbm{1}_{B''}$ is the identity on $\mathcal{H}_{B''}$ and for some $\ket{\xi_{B''E}}$, the state $\ket{\psi_{ABE}}$ is given by
\begin{eqnarray}\label{Theo1.2}
(\I_{AE}\otimes U_B)\ket{\psi_{ABE}}=\ket{\psi(\boldsymbol{\alpha})}_{AB'}
\otimes\ket{\xi_{B''E}}.
\end{eqnarray}
\end{thm}

It should be stressed that since we consider 
the adversarial scenario, we do not assume the state shared by Alice and Bob to be pure. Instead, we consider an arbitrary mixed state represented by a purification $\ket{\psi_{ABE}}$. Moreover, we do not take advantage of the assumption that Bob's measurements are projective; we infer this from maximal violation of our inequalities. 


{\it{Certification of all rank-one extremal POVM's.---}}Let us now proceed to another result, which is a consequence of Theorem \ref{Theo1}. Based on it, we can 
design a simple scheme for 1SDI certification of any extremal rank-one POVM.
%
%
Recall that a POVM $\{\mathcal{I}_b\}$, where $b$ denotes its outcomes, is called extremal if it cannot be decomposed as a convex mixture of other POVMs \cite{APP05}; moreover, a rank-one POVM is one for which the measurement operators are rank-one: $\mathcal{I}_b=\lambda_b\ket{\mu}\!\bra{\mu}$ with $\ket{\mu}\in\mathbbm{C}^d$. 

Let us go back to the 1SDI set-up and slightly modify it by adding a third measurement on the untrusted side, which is a POVM with $d^2-$outcomes, denoted by $\{R_b\}$. Moreover, we now let Alice gather statistics corresponding to $d^2$ observables constructed from the operators $X_d^iZ_d^j$ for $i,j=0,1,\ldots,d-1$ \cite{footnote1}. Notice that they contain $X_d$ and $Z_d$ used in our inequalities (\ref{Steineq1}).
Now, the maximal violation of (\ref{Steineq1}) by correlations obtained when Alice and Bob measure, respectively, $X_d$ and $Z_d$, and $B_0$ and $B_1$ are used to certify the state $\ket{\psi(\boldsymbol{\alpha})}$ shared among them (see Theorem \ref{Theo1}). Then, the correlations observed for the remaining observables of Alice and Bob's third measurement are exploited to certify the extremal POVM $\{R_b\}$.  
Indeed, generalizing the results of \cite{random1}, in Appendix \ref{App C} we prove the following fact.
\begin{thm}\label{Theo3}  
If for any $i,j=0,\ldots,d-1$ the identities 
%
%
\begin{eqnarray}\label{Vitax}
\langle X_d^iZ_d^j\otimes R_b\otimes\I_E\rangle_{\ket{\psi_{ABE}}}=\langle X_d^iZ_d^j\otimes \mathcal{I}_b\rangle_{\ket{\psi(\boldsymbol{\alpha})}}
\end{eqnarray}
hold true for the correlations obtained from the POVM $\{R_b\}_b$ while measured on 
the certified state $U_B\ket{\psi_{ABE}}=\ket{\psi(\boldsymbol{\alpha})}\otimes\ket{\xi_{B''E}}$ for some unitary transformation $U_B:\mathcal{H}_B\rightarrow\mathcal{H}_B$
and those obtained from an extremal reference POVM $\{\mathcal{I}_b\}$ with the 
ideal state $\ket{\psi(\boldsymbol{\alpha})}$, then
the elements of the POVM are given by $U_BR_bU_B^{\dagger}=\mathcal{I}_b\otimes\I_{B''}$ for all $b$.
\end{thm}

{\it{Certification of maximal randomness.---}} An important consequence of Theorem \ref{Theo3} is that one can certify $2\log_2d$ bits of randomness on the untrusted side. This is the maximum amount of randomness that can be extracted from a $d$ dimensional quantum system by performing a local measurement on it. Indeed, by virtue of Theorem \ref{Theo1}, maximal violation of our inequality for some $\boldsymbol{\alpha}$ implies that 
the joint state $\ket{\psi_{ABE}}$, after an application of a suitable local unitary 
$U_B$, takes the form (\ref{Theo1.2}). Next, Theorem \ref{Theo3} implies that 
under the same unitary Bob's additional measurement $\{R_b\}$ is $R_b=\mathcal{I}_b\otimes \mathbbm{1}$. Plugging all this into Eq. (\ref{LGpr}), and fixing $y=2$, one obtains
\begin{equation}\label{G}
    G(y=2,\vec{p})=\sup_{\sigma_E,Z}\sum_{b}\Tr[\mathcal{I}_b\rho_{B'}(\boldsymbol{\alpha})]
    \Tr[Z_b\sigma_E],
\end{equation}
where the optimization over all Eve's strategies reduces to one over the states $\sigma_E=\Tr_{B''}\proj{\xi_{B''E}}$ and Eve's measurement $Z$, and $\rho_{B'}(\boldsymbol{\alpha})=\Tr_A\proj{{\psi(\boldsymbol{\alpha})}}_{AB'}$. It is direct to see that 
for any extremal POVM $\{\mathcal{I}_b\}$ such that for any $b$,
\begin{equation}\label{ElCulo}
    \Tr[\mathcal{I}_b\rho_B(\boldsymbol{\alpha})]=1/d^2,
\end{equation}
the guessing probability (\ref{G}) becomes $G(y=2,\vec{p})=1/d^2$. 
Thus, the maximal violation of our inequality (\ref{Steineq1}) together 
with conditions (\ref{Vitax}) provide a way of certifying the maximal number $2\log_2d$
of random bits from Bob's POVM using any pure entangled state provided that there exists extremal POVM's $\{\mathcal{I}_b\}$ that satisfy (\ref{ElCulo}). An exemplary construction of extremal $d^2$-outcome POVMs obeying the condition (\ref{ElCulo}) for any $b$ and $\rho_{B'}=\I/d$ (which corresponds to the maximally entangled state of two qudits) was found in Ref. \cite{APP05}, and for completeness is presented in Appendix \ref{App D}. There, for particular cases $d=3,\ldots,6$ we also provide constructions of extremal POVM's obeying the condition (\ref{ElCulo}) for any $b$ when the certified state is non-maximally entangled (App. \ref{App D}).

{\textit{Extended Bell scenario.---}}Here we show how to extend the above self-testing results 
to a fully device-independent certification method of all two-qutrit bipartite entangled states, which is alternative to that of \cite{Projection}. Our scheme is inspired by Refs. \cite{Joe2018,zhao2021deviceindependent} and consist in a simultaneous use of our steering inequality (\ref{Steineq1}) and the Bell inequality of Ref. \cite{KST+19} (see Appendix \ref{AppF} for details). 
  \begin{figure}[h!]
    \centering
    \includegraphics[scale=0.22]{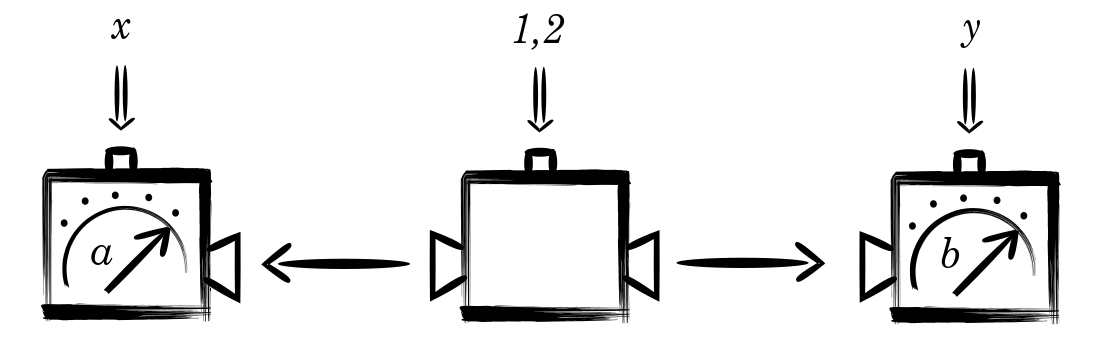}
    \caption{\textbf{Extended Bell scenario.} It consists of two parties, Alice and Bob, having access to untrusted measuring devices that measure three-outcome observables $A_x$ $(x=0,1,2)$ and $B_y$ $(y=0,\ldots,4)$, respectively, and a preparation device producing two states $\rho^i_{AB}$. First, Alice and Bob perform a Bell test on $\rho^1_{AB}$ with measurements $x,y=0,1,2$ to device-independently certify Alice's observables via violation of the Bell inequality of Ref. \cite{KST+19}. Second, the preparation device is set to produce $\rho^2_{AB}$, on which the parties measure the already certified $A_0,A_1$, and $B_3,B_4$. Maximal violation of a steering inequality (\ref{Steineq1}) for some $\boldsymbol{\alpha}$ allows them to conclude that $\ket{\psi_2}_{AB}$ is equivalent to $\ket{\psi(\boldsymbol{\alpha})}$.}
    \label{fig}
\end{figure}

Let us consider a device-independent scenario with Alice and Bob having access to untrusted measuring devices and Charlie possessing a preparation device $\mathcal{P}$ that prepares two different states $\rho^i_{AB}$ that are distributed to Alice and Bob; we denote their purifications by $\ket{\psi_{ABE}^i}$  $(i=1,2)$ (see Fig. \ref{fig}). Alice's and Bob's measuring devices have now three and five inputs, respectively. All the parties can freely choose their inputs. We additionally assume that Alice's and Bob's measuring devices are independent of the preparation choice $p$.
%
%
%

After the experiment is complete, Alice, Bob and Charlie compare their inputs along with the outcomes to reproduce the joint probability distributions $\{p(a,b|x,y,p)\}$. 
Our scheme consists of two parts. First the parties consider the correlations corresponding to the first preparation $\ket{\psi_{ABE}^1}$ and the measurements $A_x$ and $B_y$ $(x,y=0,1,2)$ to device-independently certify Alice's measurements. Indeed, as proven in Ref. \cite{KST+19}, the observation of maximal violation of the Bell inequality introduced there allows both parties to conclude that $\mathcal{H}_A=\mathbbm{C}^3\otimes\mathcal{H}_{A''}$ for some finite-dimensional Hilbert space $\mathcal{H}_{A''}$ and that there is a unitary on Alice's Hilbert space $U_A$ such that 
\begin{eqnarray}\label{Ai}
    U_A\, A_0\, U_A^{\dag}&=&Z_3\otimes \mathbbm{1}_{A'},\nonumber\\
    U_A\, A_1\, U_A^{\dag}&=&X_3\otimes Q_{A''}+X_3^{T}\otimes Q_{A''}^{\perp},
\end{eqnarray}
%
where $T$ denotes the transposition in the standard basis, $\mathbbm{1}_{A''}$ is the identity on $\mathcal{H}_{A''}$, 
and $Q_{A''},\, Q_{A''}^{\perp}$ are two orthogonal 
projections such that $Q_{A''}+Q_{A''}^{\perp}=\mathbbm{1}_{A''}$.
%

%
%
Next, the parties consider correlations resulting from the second preparation $\ket{\psi_{ABE}^2}$ where Alice measures the observables $A_0$ and $A_1$ which are now known to be of the form in (\ref{Ai}) and Bob performs the remaining measurements $B_3$ and $B_4$. If these correlations maximally violate our steering inequality (\ref{Steineq1}) for some $\boldsymbol{\alpha}$, then the parties conclude that, up to certain equivalences, the state $\ket{\psi_{ABE}^2}$ equals $\ket{\psi(\boldsymbol{\alpha})}$. Precisely, they infer that $\mathcal{H}_B=\mathbbm{C}^3\otimes\mathcal{H}_{B''}$ for some finite-dimensional $\mathcal{H}_{B''}$ and that there exists a unitary $U_B$ operation acting on $\mathcal{H}_B$ such that for some $\ket{\eta_{A''B''E}}\in\mathcal{H}_{A''}\otimes \mathcal{H}_{B''}\otimes \mathcal{H}_{E}$ (see Appendix \ref{AppF}),
\begin{equation}
    U_A\otimes U_B\otimes\I_E\ket{\psi_{ABE}^2}=\ket{\psi(\boldsymbol{\alpha})}\otimes\ket{\eta_{A''B''E}}.
\end{equation}
It should be stressed here that the fact Alice's observables can be certified with the aid of the Bell inequality of Ref. \cite{KST+19} is essential here because it enables
using our steering inequalities for certification of $\ket{\psi(\boldsymbol{\alpha})}$. In fact, it is straightforward to verify that, even if initially designed for the $X_d$ and $Z_d$ observables on the trusted side, the inequality (\ref{Steineq1}) remains  nontrivial if these observables are replaced by $A_i$ given in Eq. (\ref{Ai}). In particular, its classical value is strictly lower than the maximal quantum value for any $\boldsymbol{\alpha}$, and, at the same time, its maximal quantum value remains the same.

{\it{Conclusions and outlook.---}}In this work, we constructed a class of steering inequalities maximally violated by every entangled bipartite state. 
Our inequalities are more experimentally friendly, since they require performing only two measurements, as compered to those constructed in Ref. \cite{Dani2018} that are based on $d+1$ measurements. We then provided a method of certification of quantum realisations that maximally violate our steering inequalities. Contrary to the previous approach \cite{SBK20}, which is a direct adaptation of the 'projection' method of Ref. \cite{Projection} to the steering scenario, our scheme uses only two genuinely $d$-outcome measurements per site. This might facilitate the implementation of our method in practical scenarios. Using the self-testing results, we also introduced a simple scheme for certification of any rank-one extremal POVM, which is used to show that the maximal amount of $2\log_2d$ bits of randomness can be generated using maximally entangled state of local dimension $d$ in the 1SDI scenario as well as some partially entangled states for $d=3,4,5,6$, extending at the same time the results of  Ref. \cite{random1}.

Several follow-up questions arise from our work. First, it would be interesting to explore whether, by building on our construction, one can design Bell inequalities which are maximally violated by any entangled bipartite state and two measurements per site. Then, a related question is whether one can generalize our extended Bell scenario to other situations, such as, for instance, quantum states of higher local dimension or multipartite quantum states. 

As for randomness generation, one possibility is to find extremal POVM's of maximal number of outcomes that would result in optimal randomness generation using any pure bipartite, even arbitrarily little entangled states. Another one, certainly more challenging, is to attack the open and highly nontrivial problem of designing a DI strategy for certification of maximal randomness in quantum systems of arbitrary local dimension. A possible solution here would be to again combine our one-sided device-independent scheme with a device-independent method that allows for self-testing of measurements on Alice's side. In the particular case of $d=3$, a scheme of this type that allows for certification of $2\log_23$ bits of local randomness from any entangled two-qutrit state of Schmidt rank three can be proposed \cite{JJSR} (see also Ref. \cite{Tavakoli2021}).

\section*{Acknowledgements}
This work is supported by 
%
%
the Polish National Science Centre through the SONATA BIS project No. 2019/34/E/ST2/00369
and SONATINA project No. 2020/36/C/ST2/00592. 



\providecommand{\noopsort}[1]{}\providecommand{\singleletter}[1]{#1}%

\onecolumngrid

\appendix

\setcounter{thm}{0}

\section{On $d$-outcome quantum measurements in the observable picture}
\label{App0}

Consider a quantum measurements $N=\{N_a\}$ defined by positive-semidefinite operators $N_a$ 
acting on some finite-dimensional Hilbert space $\mathcal{H}$ that sum up to the identity on $\mathcal{H}$, $\sum_{a}N_a=\mathbbm{1}_{\mathcal{H}}$. If they are additionally pairwise
orthogonal projections, i.e., $N_aN_b=\delta_{ab}N_a$ for any choice of $a$ and $b$, then $N$ is a projective measurement; otherwise, we call it generalized measurement or POVM. 

Let us now introduce another set of operators acting on $\mathcal{H}$
that are constructed from $N_a$ via the discrete Fourier as
\begin{equation}\label{Fourier}
B_k=\sum_{a}\omega^{ka}N_a \qquad (k=0,\ldots,d-1),
\end{equation}
where, as before, $\omega=\mathrm{exp}(2\pi\mathbbm{i}/d)$. 
Due to the fact that the Fourier transform is invertible, the $B_k$ operators
uniquely identify the measurement $N$; in particular, one has
\begin{equation}
    N_a=\frac{1}{d}\sum_{k}\omega^{-ak}B_k.
\end{equation}

Let us provide a few properties of $B_k$. First, one finds that $B_0=\mathbbm{1}_{\mathcal{H}}$.
Then, from the fact that $\omega^{d-k}=\omega^{-k}=(\omega^{k})^{*}$ $(k=0,\ldots,d-1)$ it follows that 
\begin{equation}
B_{-k}=B_{d-k}=B_k^{\dagger}    
\end{equation}
for any $k$. Third, one also sees that (see Ref. \cite{KST+19} for a proof)
\begin{equation}
    B_k^{\dagger}B_k\leq \mathbbm{1}\qquad (k=0,\ldots,d-1).
\end{equation}

Let us finally discuss the particular case of $N$ being projective, in which 
case $N_a$ are mutually orthogonal projectors. Then, by the very definition (\ref{Fourier}), each $B_k$ is unitary and satisfies $B_k^d=\mathbbm{1}_{\mathcal{H}}$. Moreover, $B_k=B_1^{k}$, 
where $k$ stands for the operator power. In such a case we can simply denote $B_k=B^k$ with $B\equiv B_1$. Thus, a projective measurement $N$ is represented by 
a unitary observable $B$ and its powers.

\section{The steering inequalities and their characterization}
\label{App A}



\linespread{1.1}

This supplementary material provides detailed proofs of all results presented in the article's main body.  
For the sake of completeness of this document, we will first recall the construction of steering inequalities that we use in the article.

We consider a general bipartite entangled state in  
$\mathbbm{C}^d\otimes \mathbbm{C}^d$,
which in its Schmidt decomposition can be expressed as
\begin{eqnarray}\label{theStateApp}
\ket{\psi(\boldsymbol{\alpha})}_{AB}=\sum_{i=0}^{d-1}\alpha_{i}\ket{i}_A\ket{i}_B,
\end{eqnarray}
where $\boldsymbol{\alpha}$ is a vector composed of the Schmidt coefficients $\alpha_i$
satisfying $\sum_{i=0}^{d-1}\alpha_i^2=1$.
The local bases of \eqnref{theStateApp} are assumed to be the computational basis of $\mathbbm{C}^d$, moreover $\alpha_i>0$ for every $i$. Now, in the observable picture we define a general class of steering inequalities that are maximally violated by any two-qudit pure entangled state 
\begin{equation}\label{Stefn1}
\mathcal{B}_d(\boldsymbol{\alpha})=\sum_{k=1}^{d-1}\left\langle A_0^{k}\otimes B_{k|0} + 
\gamma(\boldsymbol{\alpha}) A_1^{k}\otimes B_{k|1}+\delta_k(\boldsymbol{\alpha})A_0^{k}\right\rangle\leq \beta_L(\boldsymbol{\alpha}),
\end{equation}
where $\gamma(\boldsymbol{\alpha})$ and $\delta_k(\boldsymbol{\alpha})$ are coefficients given by
\begin{eqnarray}\label{value1}
\gamma(\boldsymbol{\alpha})=d\left({\sum_{\substack{i,j=0\\i\ne j}}^{d-1}\frac{\alpha_i}{\alpha_j}}\right)^{-1},\qquad \delta_k(\boldsymbol{\alpha})=-\frac{\gamma(\boldsymbol{\alpha})}{d}\sum_{\substack{i,j=0\\i\ne j}}^{d-1}\frac{\alpha_i}{\alpha_j}\omega^{k(d-j)}.
\end{eqnarray}
In our scenario, Alice is considered to be trusted (or fully characterised), and her measurements are  $A_0=Z_d$ and $A_1=X_d$, where
\begin{equation}
    Z_d=\sum_{i=0}^{d-1}\omega^{i}\ket{i}\!\bra{i},\qquad X_d=\sum_{i=0}^{d-1}\ket{i+1}\!\bra{i}.
\end{equation}
Let us observe that $\mathcal{B}_d(\boldsymbol{\alpha})$ is real for any choice of Bob's observables and the coefficients $\alpha_i$. This is a consequence of the fact that $B_i$ are unitary, which implies the following property of Bob's measurements $B_{d-k|i}=B_{k|i}^{\dagger}$ and coefficients $\delta_{d-k}(\boldsymbol{\alpha})=\delta_k^*(\boldsymbol{\alpha})$.

Below we prove the classical or local bound of the \eqref{Stefn1} to be $\beta_L(\boldsymbol{\alpha}) < d$. 
In Appendix \ref{app:SOS} we demonstrate that for Bob's observables given by $B_0=Z_d^*$ and $B_1=X_d$, the quantum value of $\mathcal{B}_d(\boldsymbol{\alpha})$ is $d$. This fact implies that inequality \eqref{Stefn1} in nontrivial for any choice of $\alpha_i>0$, that is $\beta_L(\boldsymbol{\alpha})<\beta_Q(\boldsymbol{\alpha})$.

\subsection{Classical bound}
\label{app:ClassicalBound}

To determine the classical bound $\beta_L(\boldsymbol{\alpha})$ of our steering inequalities (\ref{Stefn1}), let us express the functional $\mathcal{B}_d(\boldsymbol{\alpha})$ in the probability picture as
\begin{eqnarray}
\mathcal{B}_d(\boldsymbol{\alpha})=d\sum_{a,b=0}^{d-1}c_{a,b}p(a,b|0,0)+\gamma(\boldsymbol{\alpha})\left(d\sum_{a,b=0}^{d-1}c_{a,b}p(a,b|1,1)-\sum_{\substack{i,a=0\\i\ne a}}^{d-1}\alpha_i\frac{p(a|0)}{\alpha_a}\right)-1-\gamma(\boldsymbol{\alpha})-\delta_0(\boldsymbol{\alpha}),
\end{eqnarray}
where $c_{a,b}=1$ for $a\oplus_d b=0$ and $0$ otherwise, where $\oplus_d$ stands for addition modulo $d$. Note from Eq. \eqref{value1} that $\delta_0(\boldsymbol{\alpha})=-1$, which implies that
\begin{eqnarray}
\mathcal{B}_d(\boldsymbol{\alpha})=d\sum_{a,b=0}^{d-1}c_{a,b}p(a,b|0,0)+\gamma(\boldsymbol{\alpha})\left(d\sum_{a,b=0}^{d-1}c_{a,b}p(a,b|1,1)-\sum_{\substack{i=0}}^{d-1}\alpha_i\sum_{a=0}^{d-1}\frac{p(a|0)}{\alpha_a}\right).
\end{eqnarray}
Let us now consider an LHS model 
\begin{equation}
    p(a,b|x,y)=\sum_{\lambda\in\Lambda}p(\lambda)p(a|x,\rho_{\lambda})p(b|y,\lambda),
\end{equation}
where $\Lambda$ is some set of the hidden variables $\lambda$ distributed between Alice and Bob according to probability distribution $p(\lambda)$, whereas $p(a|x,\lambda)$ and $p(b|y,\lambda)$ are local probability distributions; in particular Alice's distributions are quantum and depend
on the hidden states $\rho_{\lambda}$. For such a  model our steering functional rewrites as
\begin{eqnarray}\label{steeopprob}
\mathcal{B}_d(\boldsymbol{\alpha}) = d \sum^{d-1}_{\substack{a=0}}\sum_{\lambda\in\Lambda} \ p(\lambda)p(a|0,\rho_\lambda)p(d-a|0,\lambda)+\gamma(\boldsymbol{\alpha})\left(d \sum^{d-1}_{\substack{b=0}}\sum_{\lambda\in\Lambda}  \ p(\lambda)p(a|1,\rho_\lambda)p(d-a|1,\lambda)-\sum_i\alpha_i\sum_{a}\frac{p(a|0,\rho_A)}{\alpha_a}\right),\nonumber\\
\end{eqnarray}
where in the last term we have used the fact that Alice's local distributions 
are quantum and have substituted 
\begin{equation}
    \rho_A=\sum_{\lambda}p(\lambda)\rho_{\lambda}.
\end{equation}
%
We focus on the first two terms in Eq. (\ref{steeopprob}) and notice that 
they can be bounded from above in the following way
\begin{eqnarray}
    \sum_{a=0}^{d-1}\sum_{\lambda\in\Lambda}p(\lambda)p(a|y,\rho_{\lambda})p(d-a|y,\lambda)&\leq& \sum_{\lambda\in\Lambda}p(\lambda)\max_{a}\{p(a|y,\rho_{\lambda})\},\nonumber\\
    &\leq& \max_{\ket\psi\in\mathbbm{C}^d}\max_{a}\{p(a|y,\ket{\psi})\},
\end{eqnarray}
where $y=0,1$ and to obtain the first inequality we used the fact that $p(a|y,\lambda)$ are normalized for any $y$ and $\lambda$, and the maximization over pure states in the second inequality follows from the fact that Alice's probability distributions are linear functions of the state. As a result our expression (\ref{steeopprob}) can be bounded as
\begin{equation}
    \mathcal{B}_d(\boldsymbol{\alpha})\leq \max_{\ket{\psi}\in\mathbbm{C}^d}\left[d\max_{a}\{p(a|0,\ket{\psi})\}+\gamma(\boldsymbol{\alpha}) \left(d\max_{a}\{p(a|1,\ket{\psi})\}-\sum_{i=0}^{d-1}\alpha_i\sum_{a=0}^{d-1}\frac{1}{\alpha_a}p(a|0,\ket{\psi})\right)\right],
\end{equation}
where we used the fact that by including the last term in Eq. (\ref{steeopprob})
in the maximization over $\ket{\psi}$ the whole expression on the right-hand side of
(\ref{steeopprob}) cannot decrease.

By expanding the state $\ket{\psi}$ in the computational basis of $\mathbbm{C}^d$, 
$\ket{\psi}=\sum_{i}\eta_i\ket{i}$,
and by employing the explicit forms of Alice's observables, that is, $A_0=Z_{d}$
and $A_1=X_d$, we can rewrite the above expression in the following form
\begin{equation}\label{Leon}
    \mathcal{B}_d(\boldsymbol{\alpha})\leq \max_{\substack{|\eta_{0}|,\ldots,|\eta_{d-1}|\\|\eta_0|^2+\ldots+|\eta_{d-1}|^2=1}}\left\{d\max_{a}\{|\eta_a|^2\}+\gamma(\boldsymbol{\alpha}) \left[\left(\sum_{i=0}^{d-1}|\eta_i|\right)^2-\sum_{i=0}^{d-1}\alpha_i\sum_{a=0}^{d-1}\frac{|\eta_a|^2}{\alpha_a}\right]\right\}.
\end{equation}
Our aim now is to show that the expression on the right-hand side
is strictly less than $d$ for any set of positive $\alpha_i$ such that 
$\alpha_0^2+\ldots+\alpha_{d-1}^2=1$. To this end, we first show that 
the expression in the square brackets never exceeds zero. Using the 
the fact that $\alpha_i>0$ for every $i$, we can write
\begin{equation}\label{Asturias}
    \left(\sum_{i=0}^{d-1}|\eta_i|\right)^2=\left(\sum_{i=0}^{d-1}\sqrt{\alpha_i}\,\frac{|\eta_i|}{\sqrt{\alpha_i}}\right)^2\leq \sum_{i=0}^{d-1}\alpha_i\sum_{j=0}^{d-1}\frac{|\eta_i|^2}{\alpha_i},
\end{equation}
where we used the Cauchy-Schwarz inequality to obtain the last expression. After applying this inequality to (\ref{Leon}) we immediately conclude that $\mathcal{B}_d(\boldsymbol{\alpha})\leq d$. 
Then, it follows from (\ref{Leon}) that the case $\mathcal{B}_d(\boldsymbol{\alpha})=d$ is possible iff
\begin{equation}\label{OtherCond}
 \max_a\{|\eta_a|^2\}=1   
\end{equation}
as well as the expression in the square brackets in (\ref{Leon}) vanishes. The latter holds true iff 
both vectors in (\ref{Asturias}) are parallel, that is, $\alpha_i=\lambda |\eta_i|$
for each $i$ and some real coefficient $\lambda$; actually, due to the fact that 
$\sum_i\alpha_i^2=\sum_i|\eta_i|^2=1$, $\lambda=1$, and therefore $\alpha_i=|\eta_i|$. 
However, since $\alpha_i<1$ for any $i$, this contradicts the condition (\ref{OtherCond}), implying that $\mathcal{B}_d(\boldsymbol{\alpha})<d$ for any set of positive $\alpha_i$ such that $\alpha_0^2+\ldots+\alpha_{d-1}^2=1$. In other words, the steering inequality 
(\ref{Stefn1}) is non-trivial for any pure entangled state $\ket{\psi_{\mathrm{ideal}}}$ 
of Schmidt rank $d$.

\section{The maximal quantum value} 
\label{app:SOS}

In this section we demonstrate that the maximal quantum value of the steering functional $\mathcal{B}_d(\boldsymbol{\alpha})$ given in Eq. \eqref{Stefn1} amounts to $\beta_Q(\boldsymbol{\alpha})=d$. For this purpose we prove the following theorem.
\begin{thm}
For any set of positive and normalized coefficient $\alpha_i$,
the maximal quantum value of $\mathcal{B}_d(\boldsymbol{\alpha})$ is 
$\beta_Q(\boldsymbol{\alpha})=d$ and obeys $\beta_Q(\boldsymbol{\alpha})>\beta_L(\boldsymbol{\alpha})$.
\end{thm}
\begin{proof}Let us introduce a steering operator corresponding to 
the inequality (\ref{Stefn1}), 
\begin{equation}\label{Porto}
    \hat{\mathcal{B}}_d(\boldsymbol{\alpha})=\sum_{k=1}^{d-1}\left(A_0^k\otimes B_{k|0}+\gamma(\boldsymbol{\alpha}) A_1^k\otimes B_{k|1}+\delta_k(\boldsymbol{\alpha})A_0^k\right),
\end{equation}
where, as before, $A_0=Z_d$ and $A_1=X_d$, while $B_i$ are arbitrary $d$-outcome unitary observables measured by Bob, and prove that 
\begin{equation}\label{Lizbon}
    \max_{\ket{\psi}}\langle\psi|\hat{\mathcal{B}}_d(\boldsymbol{\alpha})|\psi\rangle\leq d.
\end{equation}
For this purpose we rewrite the operator $\hat{\mathcal{B}}_d(\boldsymbol{\alpha})$ as
\begin{equation}\label{rozklad}
    \hat{\mathcal{B}}_d(\boldsymbol{\alpha})=\sum_{k=1}^{d-1}A_0^k\otimes B_{k|0}+S(\boldsymbol{\alpha})
\end{equation}
with 
\begin{equation}\label{S1}
    S(\boldsymbol{\alpha})=\sum_{k=1}^{d-1}\left[\gamma(\boldsymbol{\alpha}) A_1^k\otimes B_{k|1}+\delta_k(\boldsymbol{\alpha}) A_0^k\right].
\end{equation}
Since $A_0$ is unitary and $B_{k|0}^{\dagger}B_{k|0}\leq \mathbbm{1}$ for any $k$, it follows that
\begin{equation}\label{sat1}
|\langle\psi| A^k_0\otimes B_{k|0}|\psi\rangle|\leq 1,  
\end{equation}
for any $\ket{\psi}$ and any $k=1,\ldots,d-1$.
Moreover, due to the fact that $A_0^{d-k}=(A_0^k)^{\dagger}$ and $B_{d-k|0}=B_{k|0}^{\dagger}$, 
the above implies that 
\begin{eqnarray}\label{Guimerais}
    \max_{\ket{\psi}}\langle\psi|\hat{\mathcal{B}}_d(\boldsymbol{\alpha})|\psi\rangle&\leq& 
    d-1+\max_{\ket{\psi}}\langle\psi|S(\boldsymbol{\alpha})|\psi\rangle.
\end{eqnarray}
Hence, to establish (\ref{Lizbon}) it suffices to prove for any $\ket{\psi}$ that $\langle\psi|S(\boldsymbol{\alpha})|\psi\rangle\leq 1$. For this purpose, we observe that any pure state from $\mathcal{H}_A\otimes\mathcal{H}_B$ can be represented as
\begin{equation}\label{representation}
 \ket{\psi_{AB}}=\sum_{i=0}^{d-1}\lambda_i\ket{i}_A\ket{e_i}_B,
\end{equation}
where $\lambda_i$ are nonnegative coefficients such that  $\lambda_0^2+\ldots+\lambda_{d-1}^2=1$, and $\ket{e_i}$ are some vectors from $\mathcal{H}_B$ which are not orthogonal in general. Using this representation we have
\begin{eqnarray}\label{Coimbra2}
\langle\psi|S(\boldsymbol{\alpha})|\psi\rangle&=&\sum_{k=1}^{d-1}\sum_{i,j=0}^{d-1}\left[\gamma(\boldsymbol{\alpha}) \lambda_i\lambda_j\bra{i}A_1^{k}\ket{j} \bra{e_i}B_{k|1}\ket{e_j}+\delta_k(\boldsymbol{\alpha})\lambda_i\lambda_j\bra{i}A_0^{k}\ket{j}\bra{e_i}e_j\rangle\right]\nonumber\\
&=&\sum_{k=1}^{d-1}\sum_{i=0}^{d-1}\left[\gamma(\boldsymbol{\alpha}) \lambda_i\lambda_{i-k} \bra{e_i}B_{k|1}\ket{e_{i-k}}+\delta_k(\boldsymbol{\alpha})\lambda_i^2\omega^{ik}\right],
\end{eqnarray}
where to obtain the second equality we employed the explicit forms of $A_i$.
Let us concentrate on the second term of the above expression and use the fact that 
$\delta_0(\boldsymbol{\alpha})=-1$ to rewrite it as
\begin{equation}
    \sum_{k=1}^{d-1}\sum_{i=0}^{d-1}\delta_k(\boldsymbol{\alpha})\lambda_i^2\omega^{ik}=1+\sum_{k=0}^{d-1}\sum_{i=0}^{d-1}\delta_k(\boldsymbol{\alpha})\lambda_i^2\omega^{ik}.
\end{equation}
Employing the explicit form of $\delta_k(\boldsymbol{\alpha})$ given in Eq. (\ref{value1}) 
and using the fact that 
\begin{equation}
    \sum_{k=0}^{d-1}\omega^{k(i-j)}=d\delta_{ij},
\end{equation}
we finally arrive at a real expression of the form
\begin{eqnarray}\label{Coimbra}
     \sum_{k=1}^{d-1}\sum_{i=0}^{d-1}\delta_k(\boldsymbol{\alpha})\lambda_i^2\omega^{ik}&=&1+\gamma(\boldsymbol{\alpha})-\gamma(\boldsymbol{\alpha})\sum_{i,j=0}^{d-1}\frac{\alpha_i}{\alpha_j}\lambda_j^2.
\end{eqnarray}
Now, we can substitute Eq. (\ref{Coimbra}) back to Eq. (\ref{Coimbra2}) and exploit the fact that $S(\boldsymbol{\alpha})$ is a Hermitian operator and that all $\lambda_i$ are positive to rewrite $\langle\psi|S(\boldsymbol{\alpha})|\psi\rangle$ as
\begin{equation}\label{rownanie}
 \langle\psi|S(\boldsymbol{\alpha})|\psi\rangle=1+\gamma(\boldsymbol{\alpha}) \sum_{k=0}^{d-1}\sum_{i=0}^{d-1}\lambda_i\lambda_{i-k} \mathrm{Re}\left(\bra{e_i}B_{k|1}\ket{e_{i-k}}\right)-\gamma(\boldsymbol{\alpha})\sum_{i,j=0}^{d-1}\frac{\alpha_i}{\alpha_j}\lambda_j^2.
\end{equation}
Using subsequently the facts that $\mathrm{Re}(z)\leq |z|$ for any $z\in\mathbbm{C}$ and that $B_{k|1}^{\dagger}B_{k|1}\leq \mathbbm{1}$ for any $k$, 
one obtains
\begin{equation}
 \langle\psi|S(\boldsymbol{\alpha})|\psi\rangle\leq 1+\gamma(\boldsymbol{\alpha}) \left[\left(\sum_{i=0}^{d-1}\lambda_i\right)^2 -\sum_{i=0}^{d-1}\alpha_i\sum_{j=0}^{d-1}\frac{\lambda_j^2}{\alpha_j}\right].
\end{equation}
Let us finally show that the expression within the square brackets does not exceed zero. Using the Cauchy-Schwarz inequality, one directly finds that
\begin{eqnarray}\label{CSineq}
    \left(\sum_{i=0}^{d-1}\lambda_i\right)^2&=&\left(\sum_{i=0}^{d-1}\sqrt{\alpha_i}\,\frac{\lambda_i}{\sqrt{\alpha_i}}\right)^2\nonumber\\
    &\leq & \sum_{i=0}^{d-1}\alpha_i\sum_{j=0}^{d-1}\frac{\lambda_j^2}{\alpha_j},
\end{eqnarray}
which means that 
\begin{equation}\label{sat2}
    \max_{\ket{\psi}}\langle\psi|S(\boldsymbol{\alpha})|\psi\rangle\leq 1
\end{equation}
and hence, taking into account Eq. (\ref{Guimerais}),
\begin{equation}
    \max_{\ket{\psi}}\langle\psi|\hat{\mathcal{B}}_d(\boldsymbol{\alpha})|\psi\rangle\leq d,
\end{equation}
which completes the proof.
\end{proof}

An important consequence of the above theorem is that if our steering functional 
(\ref{Stefn1}) is maximally violated by a state $\ket{\psi_{AB}}$ and Bob's observables
$B_i$, then the following conditions are satisfied
\begin{equation}
    \langle\psi| A_0^k\otimes B_{k|0}|\psi\rangle=1
\end{equation}
for any $k=1,\ldots,d-1$ as well as
\begin{equation}
    \langle\psi| S(\boldsymbol{\alpha})|\psi\rangle=1, 
\end{equation}
where $S_1(\boldsymbol{\alpha})$ is defined in Eq. (\ref{S1}). These conditions holds because the Bell operator presented in Eq. (\ref{rozklad}) consists of $d$ terms satisfying either 
(\ref{sat1}) or (\ref{sat2}). As a consequence, the above relations imply the following conditions for the state $\ket{\psi}$ as well as the observables $B_i$,
\begin{eqnarray}\label{SOS2}
A_0^{k}\otimes B_{k|0}\,\ket{\psi_{AB}}=\ket{\psi_{AB}} \qquad (k=1,\ldots,d-1),
\end{eqnarray}
and
\begin{eqnarray}\label{SOS3}
\sum_{k=1}^{d-1}\left[\gamma(\boldsymbol{\alpha}) A_1^{k}\otimes B_{k|1}+\delta_k(\boldsymbol{\alpha})A_0^{k}\right]\ket{\psi_{AB}}=\ket{\psi_{AB}}.
\end{eqnarray}
In the next section we use this relations to prove our self-testing statement.


\section{Self-testing of all bipartite entangled states}
\label{App C}

In this section, we present and prove one of our main results. First let us state that Bob's measurements $B_i$ $(i=0,1)$ act on $\mathcal{H}_B$, Alice's side is trusted and her Hilbert space is $\mathbbm{C}^d$, and $\mathcal{H}_E$ denotes the eavesdropper Hilbert space. In this scenario, we present the following theorem.
\begin{thm}\label{Theo1App} 
Assume that the steering inequality \eqref{Stefn1} with Alice's  observables $A_i$ defined to be $A_0=Z_d$, $A_1=X_d$ is maximally violated by 
$\ket{\psi}_{ABE}\in\mathbbm{C}^d\otimes\mathcal{H}_B\otimes\mathcal{H}_E$ 
and Bob's measurements $B_i$ $(i=0,1)$ acting on $\mathcal{H}_B$. Then the following statements hold true for any $d$: (i) $\mathcal{H}_B=\mathcal{H}_{B'}\otimes\mathcal{H}_{B''}$, where $\mathcal{H}_{B'} \equiv \mathbbm{C}^d$, $\mathcal{H}_{B''}$ is some finite-dimensional Hilbert space; (ii)
Bob's measurements are projective, meaning that $B_{b|i}=B_{i}^{b}$ for some
quantum observables $B_{i}$; (iii) there exists a local unitary transformation on Bob's side $U_B:\mathcal{H}_B\rightarrow\mathcal{H}_B$, such that
\begin{eqnarray}
\forall i, \quad U_B\,B_i\,U_B^{\dagger}=A_i^{*}\otimes\I_{B''}
\end{eqnarray}
where ${B''}$ denotes Bob's auxiliary system and the state   $\ket{\psi_{ABE}}$ is given by,
\begin{eqnarray}\label{Theo1.2App}
(\I_{AE}\otimes U_B)\ket{\psi_{ABE}}=\ket{\psi(\boldsymbol{\alpha})}_{AB'}
\otimes\ket{\xi_{B''E}}.
\end{eqnarray}
\end{thm}
\begin{proof}
In the following self-testing proof, we exploit
Eqs. \eqref{SOS2} and \eqref{SOS3} also taking into account the presence of an intruder, Eve, who in principle has access to the Bob's laboratory. We show in our scenario that despite Eve's complete knowledge of Bob's laboratory, we are able to certify his states and measurements.
Let us start by rewriting equations \eqref{SOS2} and \eqref{SOS3} taking into account the presence of Eve
\begin{eqnarray}\label{SOSST2}
\left(Z_d^{k}\otimes B_{k|0}\otimes\I_E\right)\ket{\psi_{ABE}}=\ket{\psi_{ABE}},
\end{eqnarray}
for any $k=1,\ldots,d-1$, and
\begin{eqnarray}\label{SOSST3}
\left[\sum_{k=1}^{d-1}\left(\gamma(\boldsymbol{\alpha}) X_d^{k}\otimes B_{k|1}+\delta_k(\boldsymbol{\alpha})Z_d^{k}\otimes\mathbbm{1}_B\right)\otimes\I_E\right]\ket{\psi_{ABE}}=\ket{\psi_{ABE}},
\end{eqnarray}
which stems from the fact that interference of Eve should not be detected by Alice or Bob. The proof is divided into two parts. In the first part, we determine Bobs' measurements for which our steering inequality \eqref{Stefn1} is maximally violated, whereas in the second part, we find the  $\ket{\psi_{ABE}}$ shared between Alice, Bob and Eve.

\textbf{\textit{Bob's measurements.}}
Let us begin by showing that the above conditions imply that Bob's measurements maximally 
violating our steering inequality are necessarily projective. First, by applying 
$Z_d^{-k}\otimes B_{-k|0}\equiv Z_d^{d-k}\otimes B_{d-k|0}$ to Eq. (\ref{SOSST2})
and using the facts that $Z_d$ is unitary as well as that $B_{-k|d}=B_{k|d}^{\dagger}$, one immediately obtains that 
\begin{equation}
    \mathbbm{1}_{AE}\otimes (B_{k|0}^{\dagger}B_{k|0})\ket{\psi_{ABE}}=\ket{\psi_{ABE}},
\end{equation}
where we have also used Eq. (\ref{SOSST2}) for $k\to d-k$.
This directly implies that 
\begin{equation}
    (B_{k|0}^{\dagger}B_{k|0})\rho_B=\rho_B,
\end{equation}
where $\rho_B=\Tr_{AE}[\proj{\psi_{ABE}}]$. Due to the fact that $\rho_B$ is non-singular, the above is satisfied iff $B_{k|0}^{\dagger}B_{k|0}=\mathbbm{1}_B$, and thus $B_{k|0}$ is unitary for any $k=0,\ldots,d-1$. This in turn implies that the measurement represented by 
$B_{k|0}$ is projective, that is, the positive semi-definite measurement operators 
$N_{b|0}$ are mutually orthogonal projectors. Moreover, $B_{k|0}$ are powers of $B_{1|0}$, 
and thus, in what follows, we will simply denote $B_{k|0}=B_0^k$, where $B_0\equiv B_{1|0}$.


Let us now move to proving that the second Bob's measurement is projective too. 
To this end, we will use the second condition (\ref{SOSST3}). In fact, if we consider
the following representation of the state $\ket{\psi_{ABE}}$ (analogous to that in Eq. (\ref{representation}))
\begin{equation}
    \ket{\psi_{ABE}}=\sum_{i}\lambda_i\ket{i}_A\ket{e_i}_{BE},
\end{equation}
where $\lambda_i>0$ and $\ket{e_i}_{BE}$ are some vectors shared by Bob and Eve, the condition that $\bra{\psi_{ABE}}S(\boldsymbol{\alpha})\ket{\psi_{ABE}}=1$ that stems from (\ref{SOSST3}), gives us [cf. Eq. (\ref{rownanie})]
\begin{equation}
\sum_{k=0}^{d-1}\sum_{i=0}^{d-1}\lambda_i\lambda_{i-k} \mathrm{Re}\left(\bra{e_i}[B_{k|1}\otimes\mathbbm{1}_E]\ket{e_{i-k}}\right)=\sum_{i,j=0}^{d-1}\frac{\alpha_i}{\alpha_j}\lambda_j^2,
\end{equation}
where for simplicity we have dropped the subscript $BE$. Using then 
the inequality (\ref{CSineq}) and after rearranging terms, we arrive at 
\begin{equation}\label{ineq1979}
    \sum_{k=0}^{d-1}\sum_{i=0}^{d-1}\lambda_i\lambda_{i-k} \left[\mathrm{Re}\left(\bra{e_i}[B_{k|1}\otimes\mathbbm{1}_E]\ket{e_{i-k}}\right)-1\right]\geq 0.
\end{equation}
Now, exploiting the fact that $B_{k|1}^{\dagger}B_{k|1}\leq \mathbbm{1}_B$ for any $k$, it is not difficult to see that 
\begin{equation}
    \mathrm{Re}\left(\bra{e_i}[B_{k|1}\otimes\mathbbm{1}_E]\ket{e_{i-k}}\right)\leq 1
\end{equation}
for any choice of $i$ and $k$. We then observe that since $\lambda_i>0$, 
for the inequality (\ref{ineq1979}) to be satisfied each term in the square brackets must vanish,
\begin{equation}
    \mathrm{Re}\left(\bra{e_i}[B_{k|1}\otimes\mathbbm{1}_E]\ket{e_{i-k}}\right)= 1.
\end{equation}
It is not difficult to convince oneself that the above holds true 
iff the following equation is satisfied
\begin{equation}
    [B_{k|1}\otimes\mathbbm{1}_E]\ket{e_{i-k}}=\ket{e_i},
\end{equation}
which further implies 
\begin{equation}
    \bra{e_{i-k}}[B_{k|1}^{\dagger}B_{k|1}\otimes \mathbbm{1}_E]\ket{e_{i-k}}=1.
\end{equation}

Since for a fixed $k$ the above is satisfied for any $i$, we arrive in fact at the following set of conditions 
\begin{equation}
    \bra{e_{i}}[B_{k|1}^{\dagger}B_{k|1}\otimes \mathbbm{1}_E]\ket{e_{i}}=1 \qquad (i=0,\ldots,d-1),
\end{equation}
which, after getting rid of Eve's subsystem, further implies 
$\Tr[B_{k|1}^{\dagger}B_{k|1}\rho_B^i]=1$, where $\rho_{B}^i=\Tr_E[\proj{e_i}_{BE}]$.
As before, the latter condition can be satisfied only if $B_{k|1}^{\dagger}B_{k|1}$ is an identity on the support of $\rho_B^i$. However, it is not difficult to see that the support of Bob's reduced density matrix $\rho_B=\Tr_{AE}[\proj{\psi}_{ABE}]$ is a sum of $\mathrm{supp}(\rho_B^i)$ for all $i$, $B_{k|1}^{\dagger}B_{k|1}$ is an identity on the whole Hilbert space of Bob, and thus $B_{k|1}$ is unitary for any $k$. As a result, the second Bob's measurement is projective and hence we can denote $B_{k|1}=B_{1}^k$, where $B_{1}\equiv B_{1|1}$.

Let us begin by rearranging terms in Eq. \eqref{SOSST3} and applying
$\I_{AE}\otimes B_1$ to it, which gives
\begin{eqnarray}\label{procond1.1}
\gamma(\boldsymbol{\alpha}) \sum_{k=1}^{d-1}\left( X_d^{k}\otimes B_{1}^{k+1}\right)\ket{\psi_{ABE}}=\left[\left(\I_A-\sum_{k=1}^{d-1}
\delta_k(\boldsymbol{\alpha})Z_d^{k}\right)\otimes B_{1}\right]\ket{\psi_{ABE}},
\end{eqnarray}
where, for simplicity, we  skipped the identity acting on Eve's subsystem. 
For clarity, we introduce the notation  
\begin{equation}
    \overline{Z}_A:=\I_A-\sum_{k=1}^{d-1}\delta_k(\boldsymbol{\alpha})Z_d^{k}.
\end{equation}
Next, we apply $B_{0}$ on Bob's system and obtain 
\begin{eqnarray}\label{procond2}
\gamma(\boldsymbol{\alpha})\sum_{k=1}^{d-1}\left( X_d^{k}\otimes B_0B_{1}^{k+1}\right)\ket{\psi_{ABE}}=(\overline{Z}_A\otimes B_0B_{1})\ket{\psi_{ABE}}.
\end{eqnarray}
Multiplying Eq. \eqref{procond1.1} by $Z_d^{-1}\otimes\I_B$ gives us
\begin{eqnarray}\label{eq:number1}
\gamma(\boldsymbol{\alpha})\sum_{k=1}^{d-1}\left( Z_d^{-1}X_d^{k}\otimes B_{1}^{k+1}\right)\ket{\psi_{ABE}}=(\overline{Z}_AZ_d^{-1}\otimes B_{1})\ket{\psi_{ABE}},
\end{eqnarray}
where we exploited the fact that $Z_d$ and $\overline{Z}_A$ commute.
 Then by using the commutation relation $Z_dX_d=\omega X_dZ_d$, we can rewrite Eq. \eqref{eq:number1}  as
\begin{eqnarray}\label{eq:comutationrelation}
\gamma(\boldsymbol{\alpha})\sum_{k=1}^{d-1}\left( \omega^{-k}X_d^{k}\otimes B_{1}^{k+1}\right)\left(Z_d^{-1}\otimes \I_B\right)\ket{\psi_{ABE}}=(\overline{Z}_AZ_d^{-1}\otimes B_{1})\ket{\psi_{ABE}},
\end{eqnarray}
where we also used the fact that $Z_d^{-1}\otimes\I$ and $\I\otimes B_1$ commute.
Finally, we apply Eq.  \eqref{SOSST2} for $k=d-1$, to write
\begin{eqnarray}\label{procond3}
\gamma(\boldsymbol{\alpha}) \sum_{k=1}^{d-1}
\left( \omega^{-k}X_d^{k}\otimes B_{1}^{k+1}B_0\right)\ket{\psi_{ABE}}=(\overline{Z}_A\otimes B_{1}B_0)\ket{\psi_{ABE}}.
\end{eqnarray}
In the next step, we subtract Eq. \eqref{procond3} multiplied by $\omega^{-1}$ from Eq. $\eqref{procond2}$, which 
leads us to
\begin{eqnarray}\label{procond4}
\gamma(\boldsymbol{\alpha})\sum_{k=1}^{d-1}\left[ X_d^{k}\otimes \left(B_{0}B_{1}^{k+1}-\omega^{-(k+1)}B_{1}^{k+1}B_{0}\right) \right]\ket{\psi_{ABE}}=[\overline{Z}_A\otimes (B_{0}B_{1}-\omega^{-1} B_{1}B_{0})]\ket{\psi_{ABE}}.
\end{eqnarray}
 Again, we take Eq. \eqref{SOSST3} and multiply it by $X_d^{-1}\otimes\I$, which gives us
\begin{eqnarray}\label{procond2.1}
\sum_{k=1}^{d-1}\left(\gamma(\boldsymbol{\alpha}) X_d^{k-1}\otimes B_{1}^{k}-\delta_k(\boldsymbol{\alpha})X_d^{-1}Z_d^{k}\otimes\mathbbm{1}_B\right)\ket{\psi_{ABE}}=(X_d^{-1}\otimes\mathbbm{1}_B)\ket{\psi_{ABE}}.
\end{eqnarray}
Then by multiplying \eqref{procond2.1} with  $\I_A\otimes B_{0}$, which commutes with $X_d\otimes\I_B$, we get
\begin{eqnarray}\label{procond5}
\sum_{k=1}^{d-1}\left(\gamma(\boldsymbol{\alpha}) X_d^{k-1}\otimes B_0B_{1}^{k}-\delta_k(\boldsymbol{\alpha})X_d^{-1}Z_d^{k-1}\otimes \I_B\right)\ket{\psi_{ABE}}=(X_d^{-1}Z_d^{-1}\otimes\mathbbm{1}_B)\ket{\psi_{ABE}},
\end{eqnarray}
where we also used Eq. \eqref{SOSST2} for $k=1$. 
Now, we multiply \eqref{procond2.1} by $Z_d^{-1}\otimes\I_B$ and get 
\begin{eqnarray}\label{eq:equation2}
\sum_{k=1}^{d-1}\left(\gamma(\boldsymbol{\alpha}) Z_d^{-1}X_d^{k-1}\otimes B_{1}^{k}-\delta_k(\boldsymbol{\alpha})Z_d^{-1}X_d^{-1}Z_d^{k}\otimes \I_B\right)\ket{\psi_{ABE}}=(Z_d^{-1}X_d^{-1}\otimes\mathbbm{1}_B)\ket{\psi_{ABE}}.
\end{eqnarray}
Again, we use the commutation relation $Z_dX_d=\omega X_dZ_d$, and \eqref{eq:equation2} can be simplified to
\begin{eqnarray}\label{procond6}
\sum_{k=1}^{d-1}\left(\gamma(\boldsymbol{\alpha}) \omega^{-(k-1)}X_d^{k-1}\otimes B_{1}^{k}B_0-\delta_k(\boldsymbol{\alpha})\omega X_d^{-1}Z_d^{k-1}\otimes \I_B\right)\ket{\psi_{ABE}}=\left(\omega X_d^{-1}Z_d^{-1}\otimes \I_B\right)\ket{\psi_{ABE}}.
\end{eqnarray}
Now we subtract Eq. \eqref{procond6} multiplied by $\omega^{-1}$ from Eq. \eqref{procond5} to get
\begin{eqnarray}
\gamma(\boldsymbol{\alpha})\sum_{k=1}^{d-1}\left[ X_d^{k-1}\otimes \left(B_{0}B_{1}^{k}-\omega^{-k}B_{1}^{k}B_{0}\right)\right]\ket{\psi_{ABE}}= 0,
\end{eqnarray}
which further can be broken down as follows
\begin{eqnarray}\label{procond7}
\gamma(\boldsymbol{\alpha})\sum_{k=2}^{d-1}\left[ X_d^{k-1}\otimes \left(B_{0}B_{1}^{k}-\omega^{-k}B_{1}^{k}B_{0}\right)\right]\ket{\psi_{ABE}}=- \gamma(\boldsymbol{\alpha})\left(B_{0}B_{1}-\omega^{-1}B_{1}B_{0}\right)\ket{\psi_{ABE}}.
\end{eqnarray}
Note that the L.H.S. of \eqref{procond4} and \eqref{procond7} are the same, which allows us to conclude that
\begin{eqnarray}
\overline{Z}_A\otimes (B_{0}B_{1}-\omega^{-1} B_{1}B_{0})\ket{\psi_{ABE}}=-\gamma(\boldsymbol{\alpha})\left( B_{0}B_{1}-\omega^{-1} B_{1}B_{0}\right)\ket{\psi_{ABE}},
\end{eqnarray}
which after simplification gives us,
\begin{eqnarray}\label{STF1}
\left[(1+\gamma(\boldsymbol{\alpha}))\I-\sum_{k=1}^{d-1}\delta_k(\boldsymbol{\alpha})Z_d^{k}\right]\otimes \left(B_{0}B_{1}-\omega^{-1} B_{1}B_{0}\right)\ket{\psi_{ABE}}=0.
\end{eqnarray}
As it is proven below (see Observation \ref{obsST1}), the matrix $[1+\gamma(\boldsymbol{\alpha})]\I-\sum_{k=1}^{d-1}\delta_k(\boldsymbol{\alpha})Z_d^{k}$ is invertible,
which allows us to rewrite Eq. \eqref{STF1} in the following form
\begin{equation}\label{eq:self-testing}
    (B_0B_1-\omega^{-1}B_1B_0)\ket{\psi_{ABE}}=0.
\end{equation}
%
%
%
Since without loss of generality, we can assume that $\rho_B=\Tr_{AE}\proj{\psi_{ABE}}$ is full-rank, Eq. \eqref{eq:self-testing}  implies the following 
commutation relation for Bob's observables
\begin{equation}
B_{0}B_{1}=\omega^{-1} B_{1}B_{0}.
\end{equation}
As it was proven in Ref. \cite{KST+19} the above relation implies that Bob's Hilbert space 
splits into a tensor product $\mathcal{H}_B=\mathbbm{C}^d\otimes\mathcal{H}_{B'}$
as well as there exists a unitary transformation
$U_B:\mathcal{H}_B\to\mathcal{H}_B$ such that
\begin{eqnarray}\label{BobsMeas}
U_B\,B_{0}\,U^{\dagger}_B= Z^{*}_{d}\otimes\I_{B'},\quad U_B\,B_{1}\,U_B^{\dagger}=X_d\otimes\I_{B'},
\end{eqnarray}
where $\mathbbm{1}_{B'}$ is the identity acting on $\mathcal{H}_{B'}$.

\textbf{\textit{The state}.}
 Having determined the form of Bob's measurements, we can now characterise the state $\ket{\psi_{ABE}}$ that maximally violates the steering inequality \eqref{Stefn1}. To this end, we will again use relations (\ref{SOS2}) and (\ref{SOS3}). Taking into account Bob's observables given in Eq. (\ref{BobsMeas}), they can be stated as
\begin{equation}\label{condZ}
    (Z_d\otimes Z_d^{\dagger})\ket{\psi'_{ABE}}=\ket{\psi_{ABE}'},
\end{equation}
and
\begin{eqnarray}\label{STstate11}
\sum_{k=1}^{d-1}\left[\gamma(\boldsymbol{\alpha}) X_d^{k}\otimes X_d^{k}+\delta_k(\boldsymbol{\alpha})Z^{k}\otimes\mathbbm{1}_{B'}\right]\ket{\psi'_{ABE}}=\ket{\psi'_{ABE}},
\end{eqnarray}
where $\ket{\psi_{ABE}'}=U_B\otimes\mathbbm{1}_{AE}\ket{\psi_{ABE}}$
and the operators from \eqref{condZ} and \eqref{STstate11} act on subsystems $A$ and $B'$. For simplicity we skip the identity 
acting on subsystems $B''$ and $E$.
Now, we decompose the state $\ket{\psi_{ABE}'}$
as follows
\begin{equation}
    \ket{\psi_{ABE}'}=\sum_{i,j=0}^{d-1}\ket{i}_A\ket{j}_{B'}\ket{\psi_{ij}}_{B''E},
\end{equation}
where $\ket{i}_A$ and $\ket{j}_{B'}$ represent the computational basis in $\mathbbm{C}^d$, whereas $\ket{\psi_{ij}}_{B''E}$ is some state in $\mathcal{H}_{B''}\otimes \mathcal{H}_E$. 
The condition (\ref{condZ}) implies that
\begin{eqnarray}
\sum_{i,j=0}^{d-1}\omega^{i-j}\ket{ij}\ket{\psi_{ij}}=\sum_{i,j=0}^{d-1}\ket{ij}\ket{\psi_{ij}},
\end{eqnarray}
which can be satisfied only if  
$\ket{\psi_{ij}}=0$ for any $i\neq j$. As a consequence, the state simplifies to 
\be \label{sf1}
\ket{\psi_{ABE}'}=\sum_{i=0}^{d-1}\ket{ii}\ket{\psi_{ii}}.
\ee 
Next, let us consider the condition \eqref{STstate11} in which we have extended the range of the sum to $k=0$ and we have used the fact  that $\delta_0(\boldsymbol{\alpha})=-1$. Thus, we get the following expression   
%
%
%
\begin{eqnarray}\label{STstate21}
\sum_{k=0}^{d-1}\left[\gamma(\boldsymbol{\alpha}) X_d^{k}\otimes X_d^{k}+\delta_k(\boldsymbol{\alpha})Z^{k}\otimes\mathbbm{1}_{B'}\right]
\ket{\psi_{ABE}'}
=\gamma(\boldsymbol{\alpha})\ket{\psi_{ABE}'}.
\end{eqnarray}
Then, exploiting the form of $\ket{\psi'_{ABE}}$ given in (\ref{sf1}), 
the above equation can be written as
\begin{eqnarray}\label{eq:secondcondition}
\gamma(\boldsymbol{\alpha})\sum_{k=0}^{d-1}\sum_{i=0}^{d-1}\ket{i+k}\ket{i+k}\ket{\psi_{ii}}+\sum_{k=0}^{d-1}\sum_{i=0}^{d-1}\omega^{ki}\delta_k(\boldsymbol{\alpha})\ket{ii}\ket{\psi_{ii}}=\gamma(\boldsymbol{\alpha})\sum_{i=0}^{d-1}\ket{ii}\ket{\psi_{ii}}.
\end{eqnarray}
The observation that for every $i$, $ \sum_{k=0}^{d-1}\ket{i+k}\ket{i+k}=\sum_{k=0}^{d-1}\ket{kk} $, allows us to write \eqref{eq:secondcondition} as follows
\begin{eqnarray}
\sqrt{d}\,\gamma(\boldsymbol{\alpha})\ket{\psi_{d}^+}\ket{\Psi}+\sum_{k=0}^{d-1}\sum_{i=0}^{d-1}\omega^{ki}\delta_k(\boldsymbol{\alpha})\ket{ii}\ket{\psi_{ii}}=\gamma(\boldsymbol{\alpha})\sum_{i=0}^{d-1}\ket{ii}\ket{\psi_{ii}},
\end{eqnarray}
where $\ket{\psi_{d}^+}$ is the maximally entangled state of two qudits and
$\ket{\Psi}=\sum_{i}\ket{\psi_{ii}}$. By projecting the first two subsystems onto $\ket{ii}$, we  obtain
\begin{equation}
    \gamma(\boldsymbol{\alpha})\ket{\Psi}+\sum_{k=0}^{d-1}\omega^{ki}\delta_{k}(\boldsymbol{\alpha})\ket{\psi_{ii}}=\gamma(\boldsymbol{\alpha})\ket{\psi_{ii}}.
\end{equation}
Then, using Eq. (\ref{value1}) and after making the appropriate calculations, we can find the explicit form of the state $ \ket{\psi_{ii}}$ i.e.
\begin{equation}
    \ket{\psi_{ii}}=\frac{\alpha_i}{\alpha_0+\ldots+\alpha_{d-1}}\ket{\Psi}.
\end{equation}
Combining the above result with Eq. (\ref{sf1})  finally gives us 
\begin{eqnarray}
    (U_B\otimes\mathbbm{1}_{AE})\ket{\psi_{ABE}}&=&\left(\sum_{m=0}^{d-1}\alpha_{i}\ket{ii}_{AB'}\right)\otimes\ket{\xi}_{B''E}
    =\ket{\psi(\boldsymbol{\alpha})}_{AB'} \otimes\ket{\xi}_{B''E},
\end{eqnarray}
where 
\begin{equation}
   \ket{\xi}_{B''E}=\frac{1}{\alpha_0+\ldots+\alpha_{d-1}}\ket{\Psi}. 
\end{equation}
This completes the proof.
\end{proof}

\begin{obs}\label{obsST1}
The matrix 
\begin{equation}\label{matrix}
\Tilde{Z}=\overline{Z}_A+\gamma(\boldsymbol{\alpha})\I=[1+\gamma(\boldsymbol{\alpha})]\I-\sum_{k=1}^{d-1}\delta_k(\boldsymbol{\alpha})Z_d^{k}
\end{equation}
with $\gamma$ and $\delta_k$ defined in Eq. (\ref{value1}) is positive and invertible.
\end{obs}
\begin{proof}
First of all we notice that the above operator is Hermitian which is a consequence of 
the facts that $A_0=Z_d$ and  $\delta_{k}^*=\delta_{d-k}$. Moreover, by using the explicit form of $\delta_k$ it is straightforward to see that the eigenvalues of (\ref{matrix}) are given by 
%
%
%
\begin{eqnarray}
\lambda_l=1+\gamma(\boldsymbol{\alpha})+\frac{\gamma(\boldsymbol{\alpha})}{d}\sum_{k=1}^{d-1}\sum_{\substack{i,j=0\\i\ne j}}^{d-1}\frac{\alpha_i}{\alpha_j}\omega^{k(l-j)} \quad \text{for}\quad l=0,\ldots,d-1,
\end{eqnarray}
where $\Tilde{Z}=\sum_l\lambda_l\ket{l}\!\bra{l}$.  After simple manipulations the above expression simplifies to
\begin{equation}
\lambda_l=\gamma(\boldsymbol{\alpha})\sum_{\substack{i=0}}^{d-1}\frac{\alpha_i}{\alpha_l}.
\end{equation}
Let us note that $\gamma(\boldsymbol{\alpha})$ and  $\alpha_i$'s  are positive for all $i$'s. From this fact, we conclude that all eigenvalues of the matrix (\ref{matrix}) are  positive, which in turn implies that the matrix is invertible. This statement completes the proof.
\end{proof}
%


\section{Certification of all rank-one extremal POVM}
\label{App D}

To certify any rank-one extremal POVM, we add to our $1$SDI set-up a third measurement on the untrusted Bob's side, which is a POVM with $d^2-$outcomes, denoted by $\{R_b\}$. In this updated $1$SDI set-up, Alice needs to perform $d^2$ number of measurements corresponding to $X_d^iZ_d^j$ for $i,j=0,1,\ldots,d-1$ and Bob needs to perform three measurements. The statistics obtained when Alice performs $X_d, Z_d$ with Bob performing his first two measurements are used to certify the state shared among them as shown in Theorem \ref{Theo1} and the rest of the statistics are used to certify the POVM.  
Here we present the proof of Theorem 3 that along with the self-testing statement
in Theorem \ref{Theo1} gives a method of certifying all rank-one extremal POVMs within 
the steering scenario. Recall that up to a local unitary operation $U_B$ on Bob's side 
any state $\ket{\psi_{ABE}}$ maximally violating 
our steering inequality takes the form
\begin{equation}\label{culo1}
    (\I_{AE}\otimes U_B)\ket{\psi_{ABE}}=\ket{\psi(\boldsymbol{\alpha})}_{AB'}\otimes \ket{\xi_{B''E}},
\end{equation}
where
\begin{equation}\label{culo2}
    \ket{\psi(\boldsymbol{\alpha})}_{AB'}=\sum_{i=0}^{d-1}\alpha_i\ket{ii}
\end{equation}
with $\alpha_i>0$ for any $i$.

\begin{thm}\label{Theo3App} Consider a POVM $\{R_b\}_b$ defined on Bob's Hilbert space $\mathcal{H}_{B}=\mathbbm{C}^d\otimes\mathcal{H}_{B''}$. If for some extremal 
POVM $\{\mathcal{I}_b\}_b$ defined on $\mathbbm{C}^d$ the following identities 
\begin{eqnarray}
\langle X^iZ^j\otimes R_b\otimes\I_E\rangle_{\ket{\psi_{ABE}}}=\langle X^iZ^j\otimes \mathcal{I}_b\rangle_{\ket{\psi(\boldsymbol{\alpha})}}
\end{eqnarray}
are fulfilled for any $i,j=0,\ldots,d-1$, where $\ket{\psi_{ABE}}$ and $\ket{\psi(\boldsymbol{\alpha})}_{AB'}$ are defined in Eqs. (\ref{culo1})
and (\ref{culo2}), respectively, then there exist a unitary transformation $U_B:\mathcal{H}_B\rightarrow\mathcal{H}_B$ such that
the elements of the POVM are given by $U_BR_bU_B^{\dagger}=\mathcal{I}_b\otimes\I_{B''}$ for all $b$.
\end{thm}

\begin{proof}
%
The proof presented here generalises the one of Ref. \cite{random1} to arbitrary dimension given that Alice is trusted. Eve's strategy implies that any POVM $\{R_b\}_b$ must reproduce the same statistics as the ideal POVM $\{\mathcal{I}_b\}_b$ when Eve is absent
\begin{eqnarray}\label{POVMST1}
\bra{\psi_{ABE}}X^iZ^j\otimes R_b\otimes\I_E\ket{\psi_{ABE}}=\bra{\psi(\boldsymbol{\alpha})}X^iZ^j\otimes \mathcal{I}_b\ket{\psi(\boldsymbol{\alpha})}\quad \forall\ b.
\end{eqnarray}

Let us first notice that any state $\ket{\psi(\boldsymbol{\alpha})}$ can be represented in 
terms of the maximally entangled state of two qudits $\ket{\phi_+^d}$ as
\begin{equation}\label{reprezentacja}
    \ket{\psi(\boldsymbol{\alpha})}=[\mathbbm{1}_A\otimes P(\boldsymbol{\alpha})]\ket{\phi_+^d},
\end{equation}
where 
\begin{equation}
    P(\boldsymbol{\alpha})=\sum_{i=0}^{d-1}\alpha_i\proj{i}
\end{equation}
and, as before, all $\alpha_i$ are assumed to be positive. Based on this representation, let us then introduce an operator basis for operators acting on $\mathbbm{C}^d$ given by
\begin{equation}\label{basis}
    W_{i,j}:=P(\boldsymbol{\alpha})^{-1}\left(X^{i}Z^{j}\right)^*P(\boldsymbol{\alpha})^{-1}.
\end{equation}
Notice that the elements of this basis are linearly independent because $P(\boldsymbol{\alpha})$ is invertible and $X^iZ^j$ are orthogonal in the Hilbert-Schmidt scalar product.

Now, exploiting the fact that $W_{i,j}$ form a basis, we can express each measurement operator $\mathcal{I}_b$ of the ideal POVM as 
\begin{equation}\label{POVMST6}
\mathcal{I}_b=\sum_{i,j=0}^{d-1}l^b_{i,j} W_{i,j},
\end{equation}
where $l_{i,j}^b$ are some in general complex coefficients. Analogously, taking
advantage of the fact that Bob's Hilbert space decomposes as $\mathcal{H}_{B}=\mathbbm{C}^d\otimes\mathcal{H}_{B''}$, we 
can represent each $R_b$ in the following way
\begin{eqnarray}\label{eq77}
U_BR_bU_B^{\dagger}=\sum_{i,j=0}^{d-1}W_{i,j}\otimes \widetilde{R}_{i,j}^{b},
\end{eqnarray}
where $\widetilde{R}_{i,j}^b$ are some matrices acting on $\mathcal{H}_{B''}$ and $U_B:\mathcal{H}_B\rightarrow\mathcal{H}_B$ is a unitary transformation such that from Eq. \eqref{culo1},
\begin{eqnarray}
(\I_{AE}\otimes U_B)\ket{\psi_{ABE}}=\ket{\psi(\boldsymbol{\alpha})}_{AB'}\otimes\ket{\xi_{B''E}}.
\end{eqnarray}
%
Substitution of (\ref{POVMST6}) into (\ref{POVMST1}) yields
\begin{eqnarray}
\bra{\psi_{ABE}}X^iZ^j\otimes R_b\otimes\I_E\ket{\psi_{ABE}}=\sum_{m,n}l^b_{m,n}\bra{\psi(\boldsymbol{\alpha})}X^{i}Z^{j}\otimes P(\boldsymbol{\alpha})^{-1}\left(X^{m}Z^{n}\right)^*P(\boldsymbol{\alpha})^{-1}\ket{\psi(\boldsymbol{\alpha})}
\end{eqnarray}
for any $b$. Using then Eq. (\ref{reprezentacja}) and the fact that 
$(Q\otimes R)\,\ket{\phi_+^d}=(\mathbbm{1}\otimes RQ^T)\ket{\phi_+^d}$ holds true for any two matrices $Q$ and $R$ as well as the fact that $X^iZ^j$ form an orthogonal basis in the Hilbert-Schmidt scalar product, one obtains
\begin{eqnarray}\label{eq76}
\bra{\psi_{ABE}}X^iZ^j\otimes R_b\otimes\I_E\ket{\psi_{ABE}}=\sum_{m,n}l^b_{m,n}\bra{\phi_+^d}X^{i}Z^{j}\otimes \left(X^{m}Z^{n}\right)^*\ket{\phi_+^d}=l^b_{i,j}.
\end{eqnarray}
Now, expanding the left hand side of \eqref{eq76} by using the above mentioned form of $R_b$ \eqref{eq77}, we have
%
%
\begin{eqnarray}
\bra{\xi_{B''E}}\widetilde{R}_{i,j}^{b}\otimes\I_E\ket{\xi_{B''E}}=l_{i,j}^b=\Tr\left(\widetilde{R}_{i,j}^{b}\sigma_{B''}\right),
\end{eqnarray}
where $\sigma_{B''}=\Tr_E(\proj{\xi_{B''E}})$. Using then the eigendecomposition of $\sigma_{B''}$, i.e., $\sigma_{B''}=\sum_kp_k\proj{k}$, we arrive at
\begin{eqnarray}\label{POVMST5}
\sum_kp_k\bra{k}\widetilde{R}_{i,j}^{b}\ket{k}=l_{i,j}^b.
\end{eqnarray}
Next, we introduce a collection of POVMs $\{\mathcal{I}_{b}^k\}_b$ 
numbered by $k$, whose measurement operators are given by 
\begin{eqnarray}
    \mathcal{I}_{b}^k&=&\Tr_{B''}\left[ \left(\I_{B'}\otimes\ket{k}\!\bra{k}_{B''}\right)R_b\right]\nonumber\\
    &=&\sum_{i,j=0}^{d-1}\bra{k}\widetilde{R}_{i,j}^{b}\ket{k}W_{i,j}.
\end{eqnarray}
Since $R_b\geq 0$ it directly follows from the above equation that $\mathcal{I}_{b}^k\geq 0$ for any $k$ and $b$. Moreover, $\sum_bR_b=\I_B$ implies that $\sum_b\mathcal{I}_{b,k}=\I_{B'}$ for any $k$. All this means that $\{\mathcal{I}_b^k\}_b$ are indeed proper quantum measurements. 

These additional POVMs, through Eq. (\ref{POVMST5}), allows us to decompose
the ideal POVM $\{\mathcal{I}_b\}$ in the following way 
\begin{eqnarray}
\mathcal{I}_{b}=\sum_kp_k\mathcal{I}_{b}^k,
\end{eqnarray}
which, given that the ideal POVM is assumed to be extremal, implies that 
\begin{eqnarray}
\forall k\qquad \mathcal{I}_{b}^k=\mathcal{I}_{b},
\end{eqnarray}
which can equivalently understood as
\begin{eqnarray}\label{eq83}
\forall k \qquad \bra{k}\widetilde{R}_{i,j}^{b}\ket{k}=l_{i,j}^b.
\end{eqnarray}

Now, we consider the following vectors from $\mathbbm{C}^d$:
\begin{eqnarray}\label{eq84}
\ket{\varphi_{a,s,t}}=\frac{1}{\sqrt{2}}\left(\ket{s}\pm\mathbbm{i}^a\ket{t}\right),
\end{eqnarray}
where $\ket{s}$ and $\ket{t}$ are vectors belonging to the eigenbasis $\{\ket{k}\}$ of $\sigma_{B''}$ such that $s\ne t$ and $a=0,1$. Let us look at the quantity
\begin{eqnarray}
\Tr_{B''}\left[\left(\I_{B'}\otimes|\varphi_{a,s,t}\rangle\!\langle \varphi_{a,s,t}|_{B''}\right)R_b\right]=\sum_{i,j}\Tr(|\varphi_{a,s,t}\rangle\!\langle \varphi_{a,s,t}|_{B''}\widetilde{R}_{i,j}^{b}) W_{i,j},
\end{eqnarray}
which with the aid of the explicit form of the vector \eqref{eq84}
can be rewritten as
\begin{eqnarray}
\Tr_{B''}\left[\left(\I_{B'}\otimes|\varphi_{a,s,t}\rangle\!\langle \varphi_{a,s,t}|_{B''}\right)R_b\right]=\mathcal{I}_b\pm
\Tr_{B''}\left[(\mathbbm{1}_{B'}\otimes L^a_{B''})R_b\right],
\end{eqnarray}
where $L_{B''}^a=(\mathbbm{i}^a/2)\left(|t\rangle\!\langle s|+(-1)^a|s\rangle\!\langle t|\right)$. Using the fact that $R_b$ is positive, implying that the left-hand side of the above 
relations is nonnegative, allows us to conclude that
\begin{eqnarray}\label{eq87}
\mathcal{I}_b\geq
\pm\Tr_{B''}\left[(\mathbbm{1}_{B'}\otimes L^a_{B''})R_b\right].
\end{eqnarray}
From Ref. \cite{APP05} it follows that the measurement operators of any rank-one extremal POVM can be written as $\mathcal{I}_b=\lambda_b\proj{\mu_b}$, where for each $b$, $\ket{\mu_b}$ is a normalized vector from $\in\mathbbm{C}^d$. As it is explicitly demonstrated at the end of the proof, the fact that $\mathcal{I}_b$ is rank one implies that the operator appearing on the right-hand side of Eq. (\ref{eq87}) must be rank one too, and, moreover, it must be of the following form
\begin{eqnarray}\label{POVMST2}
\Tr_{B''}\left[(\mathbbm{1}_{B'}\otimes L^a_{B''})R_b\right]=\lambda'_b\proj{\mu_b},
\end{eqnarray}
where $\lambda_b\geq\pm\lambda'_b$. Using then the facts that $\sum_b R_b=\I$ and that
$\Tr L_{B''}^a=0$ for any $a$, we conclude that
\begin{eqnarray}
\sum_b\Tr_{B''}\left[(\mathbbm{1}_{B'}\otimes L^a_{B''})R_b\right]=0=\sum_b\lambda_b'\proj{\mu_b}.
\end{eqnarray}
Since, $\mathcal{I}_b$ are linearly independent the above condition implies that $\lambda'_b=0$ for all $b$. Thus, from \eqref{POVMST2} we can conclude that $\Tr_{B''}\left[(\mathbbm{1}_{B'}\otimes L^a_{B''})R_b\right]=0$ for any $b$ and $a$, which 
taking into account the explicit form of $L^a_{B''}$ results in the following conditions
\begin{eqnarray}\label{POVMST3}
\left(X^iZ^j\right)^*\left(\bra{s}\widetilde{R}_{i,j}^{b}\ket{t}+\bra{t}\widetilde{R}_{i,j}^{b}\ket{s}\right)=0,
\end{eqnarray}
for $a=0$ and for $a=1$, we have
\begin{eqnarray}\label{POVMST4}
\left(X^iZ^j\right)^*\left(\bra{t}\widetilde{R}_{i,j}^{b}\ket{s}
-\bra{s}\widetilde{R}_{i,j}^{b}\ket{t}\right)=0,
\end{eqnarray}
where we used the fact that $X^iZ^j$ are linearly independent for all $i,j$. The only possible solution of equations \eqref{POVMST3} and \eqref{POVMST4} is $\bra{s}\widetilde{R}_{i,j}^{b}\ket{t}=0$ for $s\ne t$. Thus, 
from Eq. \eqref{eq83} we conclude that the POVM acting on Bob's part is given by $R_b=\mathcal{I}_b\otimes\I_{B''}$ for all $b's$.

To complete the proof let finally show that Eq. (\ref{POVMST2}) holds true for any $b$. To this end, let us pick a particular $b$. Clearly, using the corresponding vector $\ket{\mu_b}$ we can construct an orthonormal basis 
$\ket{\phi_i}$ with $i=0,\ldots,d-1$ such that $\ket{\phi_b}=\ket{\mu_b}$.
Sandwiching then inequalities \eqref{eq87} with $\ket{\mu_j}$ and $\ket{\mu_i}$, 
one obtains
\begin{eqnarray}\label{VinoTinto}
\forall_{i,j}\qquad \bra{\mu_j}\mathcal{I}_b\ket{\mu_i}\geq
\pm\bra{\mu_j}\Tr_{B''}\left[(\mathbbm{1}_{B'}\otimes L^a_{B''})R_b\right]\ket{\mu_i}.
\end{eqnarray}
For $i\neq b$ or $j\neq b$ the above gives
\begin{eqnarray}
0\geq\pm\bra{\mu_j}\Tr_{B''}\left[(\mathbbm{1}_{B'}\otimes L^a_{B''})R_b\right]\ket{\mu_i},
\end{eqnarray}
which directly implies that $\bra{\mu_j}\left[(\mathbbm{1}_{B'}\otimes L^a_{B''})R_b\right]\ket{\mu_i}=0$ for any $i\neq b$ or $j\neq b$ and simultaneously proves Eq. (\ref{POVMST2}). Moreover, for $i=j=b$, Eq. (\ref{VinoTinto}) gives 
\begin{equation}
    \lambda_b\geq \pm \bra{\mu_b}\Tr_{B''}\left[(\mathbbm{1}_{B'}\otimes L^a_{B''})R_b\right]\ket{\mu_b},
\end{equation}
which additionally imposes that $\lambda_b\geq \pm\lambda_b'$, completing the proof.
\end{proof}

\section{Construction of $d^2$-outcome  extremal POVM}
\label{App E}


Here we provide an example of extremal qudit POVM that produces locally uniformly random results from two-qudit maximally entangled state $\ket{\psi(\boldsymbol{\alpha})}_{AB}$.

It should be mentioned that such extremal POVMs were already   
studied in Ref. \cite{APP05}. In particular, it was proven there that 
the measurement operators defining $d^2$-outcome extremal POVMs must necessarily be rank-one, and, moreover, an exemplary construction thereof was given in Hilbert space of arbitrary dimension $d$. For completeness we begin our considerations by recalling this construction. To this aim, let us
define the following set of $d^2$ unitary operators
\begin{equation}
    U_{k,l}=X_d^kZ_d^l,
\end{equation}
where $k,l=0 \dots d-1$. An example of extremal $d^2$-outcome POVM $\{\mathcal{I}_{k,l}\}$ in dimension $\mathbbm{C}^d$ is then given by 
\begin{equation}
    \mathcal{I}_{k,l}:=\frac{1}{d} U_{k,l} \ket{\nu}\!\bra{\nu} U^\dag_{k,l},
\end{equation}
where $\ket{\nu}$ is any pure state from the $d-$dimensional Hilbert space that satisfies the condition $\Tr[U^\dag_{k,l} \ket{\nu} ] \ne 0 $ for all $k,l$. Noticeably, such a POVM can be used to generate $2\log_2d$ bits of randomness from the two-qudit maximally entangled state $\ket{\phi_+^d}$ since both its reduced density matrices are simply the maximally mixed state $\rho_A=\rho_B=\mathbbm{1}_d/d$, and therefore
$\Tr[  \mathcal{I}_{k,l} \rho_B]=1/d^2$.  

Now, we demonstrate the existence of extremal qudit POVMs with $d^2$ outcomes for $d=3,4,5,6$, that result in $2\log_2d$ bits of randomness from partially entangled pure states $\ket{\psi(\boldsymbol{\alpha})}_{AB}$  
for $\alpha_i\geq 1/d$ for $i=0,1,\ldots,d-2$.
Let us consider the following rank-one positive semi-definite operators: 
\begin{equation}\label{eq:povm elements}
    \mathcal{I}_{b}:=\lambda_{b} \ket{\delta_b}\!\bra{\delta_b},
\end{equation}
with $b=0, \ldots,  d^2-1$, we have
\begin{eqnarray} 
    \ket{\delta_i}&=&\ket{i},\qquad (i=0,\ldots,d-2).
\end{eqnarray}
The rest of the vectors, that is, those for $b=d-1,d,\ldots,d^2-1$ are given by
\begin{eqnarray}\label{POVMele2}
    \ket{\delta_b}=\sum_{i=0}^{d-1}\mu_i \exp{\left( \frac{2\pi \mathbbm{i}\xi_i(b-d+1)}{d^2-d+1}\right)}\ket{i},
\end{eqnarray}
where
\begin{eqnarray}
    \mu_i=\sqrt{\frac{1-\lambda_i}{(d^2-d+1)\lambda_d}}\qquad (i=0,1,\ldots,d-2),
\end{eqnarray}
and
\begin{equation}
    \quad\mu_{d-1}=\sqrt{\frac{1}{(d^2-d+1)\lambda_d}}.
\end{equation}
The $\lambda_b's$ are given by
\begin{eqnarray}
    \lambda_i=\frac{1}{d^2\alpha_i^2}\qquad (i=0,1,\ldots,d-2), 
\end{eqnarray}
and
\begin{equation}
 \lambda_{d-1}=\lambda_d=\ldots=\lambda_{d^2-1}=\frac{1}{d^2-d+1}\left(d-\sum_{i=0}^{d-2}\lambda_i\right).
\end{equation}
Finally, we provide the $\xi_i$ coefficients. Let us note here that to ensure the linear independence of the POVM elements \eqref{eq:povm elements}, the $\xi_i$ coefficients should satisfy the following property $\forall \, i\ne j, \,\, |\xi_i-\xi_j|=n$, where $n=1,\dots,d(d-1)/2$. Below we list examples of possible solutions for the $\xi_i$ coefficients  in a given dimension.  \\
For $d=3$:
%
\begin{eqnarray}
    \xi_0=0,\quad\xi_1=1\quad\text{and}\quad\xi_2=3.
\end{eqnarray}
For $d=4$:
\begin{eqnarray}
    \xi_0=0,\quad\xi_1=1,\quad \xi_2=3\quad\text{and}\quad\xi_3=9.
\end{eqnarray}
For $d=5$:
\begin{eqnarray}
    \xi_0=0,\quad\xi_1=1,\quad\xi_2=4,\quad\xi_3=14\quad\text{and}\quad\xi_4=16.
\end{eqnarray}
For $d=6$:
\begin{eqnarray}
    \xi_0=0,\quad\xi_1=1,\quad \xi_2=3,\quad \xi_3=8,
    \quad\xi_4=12\quad\text{and}\quad\xi_4=18.
\end{eqnarray}
%
 
\section{Extended Bell scenario}
\label{AppF}

This section shows how our approach to certification of all bipartite entangled states
from $\mathbbm{C}^3\otimes\mathbbm{C}^3$ based on quantum steering can be made fully device-independent by combining it with the current self-testing strategy for the maximally entangled state of two qutrits and mutually unbiased bases provided in Ref. \cite{KST+19}. 
This strategy relies on violation of a certain Bell inequality that for the particular case
of $d=3$ considered here can be stated as
\begin{equation}\label{SuperApp}
    \sum_{k=1}^2\sum_{x,y=0}^{2}\lambda_k\,\omega^{kxy}\left\langle A_{k|x}\otimes B_{k|y}\right\rangle\leq 6\sqrt{3}\cos\left(\frac{\pi}{9}\right),
\end{equation}
%
%
where $\lambda_0=1$, $\lambda_1=\mathrm{e}^{-\mathbbm{i}\pi/18},\lambda_2=\lambda_1^*$. The above inequality is maximally violated by the maximally entangled state of two qutrits $\ket{\phi_+^3}$ 
and observables $A_x, B_y$ for $x,y\in \{0,1,2\}$ acting on, respectively, $\mathcal{H}_A$ and $\mathcal{H}_B$ whose eigenvectors form mutually unbiased bases. Actually, the following statement follows directly from Ref. \cite{KST+19}.
\begin{fakt}\label{VihnoVerde}
Assume that $\ket{\psi_{ABE}}\in\mathcal{H}_A\otimes\mathcal{H}_B\otimes \mathcal{H}_{E}$ 
together with the measurements $A_x=\{A_{k|x}\}$ and $B_y=\{B_{k|y}\}$ maximally violate the inequality (\ref{SuperApp}). Then, 
$\mathcal{H}_A=\mathbbm{C}^3\otimes\mathcal{H}_{A''}$ and 
$\mathcal{H}_B=\mathbbm{C}^3\otimes\mathcal{H}_{B''}$ for some Hilbert spaces on Alice's and Bob's sides, $\mathcal{H}_{A''}$ and $\mathcal{H}_{B''}$, whose dimensions are  unknown but finite. Moreover, Alice's and Bob's measurements are projective, meaning that 
$A_{k|x}=A^{k}_x$ for some unitary $A_x$ such that $A_x^3=\mathbbm{1}$ (the same holds for Bob's measurements). Finally, there exist unitary operations 
$U_A:\mathcal{H}_A\to \mathcal{H}_A$ and $U_B:\mathcal{H}_B\to \mathcal{H}_B$
such that in particular the state and two observables of Alice, say $A_0$ and $A_1$, 
obey
\begin{equation}
    U_A\otimes U_B\otimes\I_E\ket{\psi_{ABE}}=\ket{\phi_+^3}_{A'B'}\otimes\ket{\varphi_{A''B''E}},
\end{equation}
and
\begin{eqnarray}
U_A\,A_0\,U_A^{\dagger}=Z_3\otimes \I_{A''},\qquad U_AA_1U_A^{\dagger}=X_3\otimes Q_{A''}+X_3^T\otimes Q_{A''}^{\perp},
\end{eqnarray}
where $Q_{A''}$ and $Q_{A''}^{\perp}$ are orthogonal projections 
such that $Q_{A''}+Q_{A''}^{\perp}=\mathbbm{1}_{A''}$.
\end{fakt}

Let us stress here that in Ref. \cite{KST+19} an analogous characterization was also given to the third observable of Alice and all three observables of Bob. However, since they are not used in the considerations that follow, we do not present their explicit forms.

Let us now consider a device-independent scenario with Alice and 
Bob, having access to untrusted measuring devices in their laboratories, and Charlie possessing a preparation device $\mathcal{P}$ that prepares two different, in general mixed states $\rho_{AB}^i$ $(i=1,2)$ that are distributed to Alice and Bob; we will denote purifications of these states by $\ket{\psi_{ABE}^i}$.
Alice's and Bob's measuring devices perform now, respectively, three and five measurements, which might in general be arbitrary POVM's; slightly abusing the 
notation we denote them $A_x$ $(x=0,1,2)$ and $B_y$ $(y=0,\ldots,4)$.

Our scheme consists of two steps. In the first one, Alice and Bob consider the statistics corresponding to the preparation $\ket{\psi_{ABE}^1}$ and the measurements $A_x,B_y$ $(x,y=0,1,2)$. 
If these correlations maximally violate the Bell inequality (\ref{SuperApp}), Alice can deduce through Fact \ref{VihnoVerde} that all her measurements are projective, meaning that $A_x$ $(x=0,1,2)$ are quantum observables, and that they are of the form 
\begin{eqnarray}\label{Wawel}
    A_0=U_A^{\dagger}(Z_3\otimes\mathbbm{1}_{A''})\,U_A,\qquad
    A_1=U_A^{\dagger}(X_3\otimes Q_{A''}+X_3^T\otimes Q_{A''}^{\perp})\,U_A,
\end{eqnarray}
where the auxiliary Hilbert space $\mathcal{H}_{A''}$ as well as the unitary $U_A$
remain unknown; we only know that these two objects exist and that $\mathcal{H}_{A''}$ is finite-dimensional. In this sense, maximal violation of (\ref{SuperApp}) provides only a partial characterization of Alices observables, which, nevertheless, is sufficient for the second part of our scheme which exploits the steering inequalities introduced in the present work. In fact, one can check that for this type of observables on Alice's side our steering inequalities (\ref{Stefn1}) remain nontrivial, i.e., $\beta_L(\boldsymbol{\alpha})<\beta_Q(\boldsymbol{\alpha})$ for any choice of $\boldsymbol{\alpha}$ such that $\alpha_i>0$. This is a consequence of the fact that they are constructed from $X_3$ and $Z_3$ operators.

Next, the correlations corresponding to the second preparation $\ket{\psi_{ABE}^2}$ and the measurements $A_0,A_1$ (that have just been certified to be of the form (\ref{Wawel})) and $B_3,B_4$ are considered  along with the steering inequality (\ref{Stefn1}) to verify that the second state prepared by the device $\mathcal{P}$ is, up to certain equivalences, $\ket{\psi(\boldsymbol{\alpha})}$. Specifically, the maximal violation of (\ref{Stefn1}) by 
$A_0$ and $A_1$ given in (\ref{Wawel}) and the other two measurements of Bob, 
$B_3$ and $B_4$, implies that: (i) $\mathcal{H}_{B}=\mathbbm{C}^3\otimes\mathcal{H}_{B''}$ for some finite-dimensional Hilbert space $\mathcal{H}_{B''}$; (ii) both $B_3$ and $B_4$ are quantum observables; (iii) there exists a unitary on Bob's Hilbert space $\mathcal{H}_B$ such that
\begin{equation}\label{Lego}
    U_A\otimes U_B\otimes\I_E\ket{\psi_{ABE}^2}=\ket{\psi(\boldsymbol{\alpha})}\otimes\ket{\eta_{A''B''E}}
\end{equation}
for some state $\ket{\eta_{A''B''E}}\in\mathcal{H}_{A''}\otimes \mathcal{H}_{B''}\otimes \mathcal{H}_{E}$. 

Let us present a proof of this fact. The state $\ket{\psi_{ABE}^2}$
that gives rise to the maximal violation of (\ref{Stefn1}) with Alice's observables (\ref{Wawel}) can always be written as $\ket{\psi_{ABE}^2}=(U_A^{\dagger}\otimes\mathbbm{1}_{BE})\ket{\psi'_{ABE}}$ for some $\ket{\psi'_{ABE}}$. Clearly, the latter maximally violates (\ref{Stefn1})
with Alice's observables of the following form
\begin{eqnarray}\label{Wawel2}
    A_0'=Z_3\otimes\mathbbm{1}_{A''},\qquad
    A_1'=X_3\otimes Q_{A''}+X_3^T\otimes Q_{A''}^{\perp}.
\end{eqnarray}
While we managed to get rid of the unitary operation $U_A$, 
one observes that the second Alice's observable is a combination of $X_3$
and its transposition acting on orthogonal supports. To deal with this difficulty,
we can represent $\ket{\psi'_{ABE}}$ as
\begin{equation}\label{rozklad2}
    \ket{\psi'_{ABE}}=\alpha\ket{\varphi}_{ABE}+\beta\ket{\varphi^{\perp}}_{ABE},
\end{equation}
where $\ket{\varphi}_{ABE}$ and $\ket{\varphi^{\perp}}_{ABE}$ are normalized 
projections of $\ket{\psi_{ABE}'}$ onto the subspaces corresponding to $Q_{A''}$ and $Q_{A''}^{\perp}$, respectively, that is,
\begin{equation}
    \ket{\varphi}_{ABE}=\left(\I_{A'}\otimes  \mathcal{Q}_{A''} \otimes \I_{BE} \right)\ket{\psi_{ABE}'},\qquad \ket{\varphi^{\perp}}_{ABE}=(\I_{A'}\otimes \mathcal{Q}_{A''}^{\perp} \otimes \I_{BE} )\ket{\psi_{ABE}'}, 
\end{equation}
and $\alpha$ and $\beta$ are positive numbers such that $\alpha^2+\beta^2=1$. Clearly,
$\langle\varphi|\varphi^{\perp}\rangle=0$.

%
%

Now, one directly realizes that both $\ket{\varphi}_{ABE}$ and 
$\ket{\varphi^{\perp}}_{ABE}$ violate maximally the steering inequality (\ref{Stefn1})
for the same $\boldsymbol{\alpha}$, however, the first state for the following observables on the trusted side
\begin{equation}\label{choice1}
    \widetilde{A}_0=Z_3\otimes Q,\qquad \widetilde{A}_1=X_3\otimes Q,
\end{equation}
whereas the second one for
\begin{equation}\label{choice2}
    \widetilde{A}'_0=Z_3\otimes Q^{\perp},\qquad \widetilde{A}'_1=X^{\dagger}_3\otimes Q^{\perp}.
\end{equation}
In Eqs. (\ref{choice1}) and (\ref{choice2}) we skipped the subscript 
$A''$ to make the mathematical formulas easier to read.

It is direct to check that 
\begin{equation}
    \widetilde{A}_0\,\widetilde{A}_1=\omega \widetilde{A}_1\,\widetilde{A}_0
\end{equation}
as well as
\begin{equation}
    \widetilde{A}_0'\,\widetilde{A}_1'=\omega^2 \widetilde{A}_1'\,\widetilde{A}_0'.
\end{equation}
Consequently, for both states $\ket{\varphi}_{ABE}$ and $\ket{\varphi^{\perp}}_{ABE}$ and the corresponding choices of observables on the trusted side (\ref{choice1}) and (\ref{choice2}), one can apply the reasoning of the proof of Theorem \ref{Theo1} and conclude that in both cases there exists a unitary operation $V^1_B$ and $V^2_B$ acting on local supports of both the vectors $\ket{\varphi}_{ABE}$ and $\ket{\varphi^{\perp}}_{ABE}$ such that 
\begin{equation}\label{Wedel1}
    (\mathbbm{1}_{AE}\otimes V^1_B)\ket{\varphi}_{ABE}=(\alpha_0\ket{00}+\alpha_1\ket{11}+\alpha_2\ket{22})\otimes
    \ket{\xi_{A''B''E}^1}
\end{equation}
and
\begin{equation}\label{Wedel2}
    (\mathbbm{1}_{AE}\otimes V^2_B)\ket{\varphi^{\perp}}_{ABE}=(\alpha_0\ket{00}+\alpha_1\ket{11}+\alpha_2\ket{22})
    \otimes\ket{\xi_{A''B''E}^2}
\end{equation}
for some pure states $\ket{\xi_{A''B''E}^1}$ and $\ket{\xi_{A''B''E}^2}$. To complete the proof we still need to wrap these two unitary operations $V^1_B$ and $V_B^2$ into a single one, and to reach this goal we will prove that the local supports of $\ket{\varphi}_{ABE}$ and $\ket{\varphi^{\perp}}_{ABE}$ corresponding to Bob's reduced density matrices are orthogonal. 

For this purpose, we denote $V_i=\mathrm{supp}(\sigma^i_B)$ and $V=\mathrm{supp}(\rho_B)$, where $\sigma_B^1=\Tr_{AE}[\proj{\varphi}_{ABE}]$, $\sigma_B^2=\Tr_{AE}[\proj{\varphi^{\perp}}_{ABE}]$, and $\rho_B=\Tr_{AE}[\proj{\psi'}_{ABE}]$. Moreover, let $R_i$ and $R$ be
the projectors onto $V_i$ and $V$, respectively. 
We can now consider projections of Bob's measurements
onto these supports $V_i$ and $V$, that is, 
\begin{equation}
    \widetilde{B}_{b|y}=R_1\,B_{b|y}\,R_1,\qquad \widetilde{B}'_{b|y}=R_2\,B_{b|y}\,R_2,\qquad B_{b|y}'=R\,B_{b|y}\,R.
\end{equation}
Recall that in the case $d=3$, Bob's measurements $B_3$ and $B_4$ are defined by three positive semi-definite operators $B_{b|y}\geq 0$ with $b=0,1,2$, which satisfy $B_{2|y}=B_{1|y}^{\dagger}$ and $B_{0|y}=\mathbbm{1}$ (cf. Appendix \ref{App0}).
Since both the projected measurements $\{\widetilde{B}_{b|y}\}$ and $\{\widetilde{B}'_{b|y}\}$ give rise to maximal violation of our steering inequality, it follows from Theorem \ref{Theo1.2App} that $\widetilde{B}_{b|y}$ as well as
$\widetilde{B}'_{b|y}$ are unitary on $V_1$ and $V_2$, respectively, and therefore, as before, we can represent them by quantum observables $\widetilde{B}_y$ and $\widetilde{B}_y'$.


Now, it is crucial to realize that the fact that $\widetilde{B}_{y}$
and $\widetilde{B}_y'$ are unitary implies that $B_{b|y}$ 
split into blocks acting on $V_i$ and their orthocomplements, that is,
\begin{equation}\label{Block1}
    B_{1|i}=\left(
    \begin{array}{cc}
    \widetilde{B}_{i} & 0\\
    0 & E_{i}
    \end{array}
    \right)
\end{equation}
and, simultaneously,
\begin{equation}\label{Block2}
    B_{1|i}=\left(
    \begin{array}{cc}
    \widetilde{B}'_{i}& 0\\
    0 & E'_{i}
    \end{array}
    \right),
\end{equation}
where both $E_i$ and $E_i'$ some operators
satisfying $E_iE_i^{\dagger}\leq \mathbbm{1}$ and 
$E_i'E_i'^{\dagger}\leq \mathbbm{1}$.

To prove this statement we consider a general block form of $B_{1|i}$ in terms of
the subspace $V_1$ and its orthocomplement,
\begin{equation}\label{Block3}
    B_{1|i}=\left(
    \begin{array}{cc}
    \widetilde{B}_i & C_i \\D_i &E_i
    \end{array}
    \right),
\end{equation}
where $\widetilde{B}_i$ are unitary on $V_i$ and $C_i$, $D_i$ and $E_i$ are some blocks.
Using the fact that $B_{1|i}B_{1|i}^{\dagger}\leq\I$, we arrive at the following relation
\begin{equation}
   B_{1|i}B_{1|i}^{\dagger}=\left(
    \begin{array}{cc}
    \I+C_iC_i^{\dagger} & \widetilde{B}_iD_i^{\dagger}+C_iE_i^{\dagger} \\D_i\widetilde{B}_i^{\dagger}+E_iC_i^{\dagger}&D_iD_i^{\dagger}+E_iE_i^{\dagger}
    \end{array}
    \right)\leq\left(
    \begin{array}{cc}
    \I & 0 \\0&\I
    \end{array}
    \right),
\end{equation}
which implies in particular that $\I+C_iC_i^{\dagger}\leq \mathbbm{1}$. As a result $C=0$. Similarly, using the fact that $B_{1|i}^{\dagger}B_{1|i}\leq\I$ (cf. Appendix \ref{App0}) we obtain $D=0$, and finally $E_iE_i^{\dagger}\leq\I$ as well as $E_i^{\dagger}E_i\leq\I$. In analogous manner one also derives (\ref{Block2}).

%
%
%
%
%
Using then the block forms (\ref{Block1}) and (\ref{Block2}), one immediately realizes
that
\begin{equation}\label{commutrel}
    [B_i,R_1]=[B_i,R_2]=0.
\end{equation}
On the other hand, following the reasoning of
the proof of Theorem \ref{Theo1}, one realizes that 
the projected observables $\widetilde{B}_i$ and $\widetilde{B}_i'$ 
fulfil the following identities,
\begin{equation}
    \widetilde{B}_0\widetilde{B}_1=\omega^2\widetilde{B}_1\widetilde{B}_0
\end{equation}
and
\begin{equation}
    \widetilde{B}'_0\widetilde{B}'_1=\omega\widetilde{B}'_1\widetilde{B}'_0,
\end{equation}
which by virtue of the commutation relations (\ref{commutrel})
can be rewritten as
\begin{equation}
 B_0B_1R_1=\omega^2B_1B_0R_1
\end{equation}
and
\begin{equation}
 B_0B_1R_2=\omega B_1B_0R_2.
\end{equation}
Now, we conjugate the second equation and apply it to the first one. This leads us to 
\begin{equation}
    R_2B_0^{\dagger}B_1^{\dagger}B_1B_0R_1=\omega R_2B_0^{\dagger}B_1^{\dagger}B_1B_0 R_1,
\end{equation}
which due to the fact that both $B_i$ are unitary implies that $R_2R_1=\omega R_2R_1$
meaning that $R_2R_1=0$. Thus, both subspaces $V_i$ are orthogonal; in fact, $\mathcal{H}_B=V_1\oplus V_2$. Consequently, since both unitary operations $V_B^i$
act on orthogonal subspaces they can be used to construct a single unitary operation
$U_B=V_B^1\oplus V_B^2$ for which, taking into account Eqs. (\ref{Wedel1}) and (\ref{Wedel2})
and Eq. (\ref{rozklad2}), one has
\begin{equation}
    (\mathbbm{1}_A\otimes U_B)\ket{\psi_{ABE}'}=(\alpha_0\ket{00}+\alpha_1\ket{11}+\alpha_2\ket{22})\otimes
    \ket{\eta}_{A''B''E},
\end{equation}
where $\ket{\eta}_{A''B''E}=\alpha\ket{\xi_{A''B''E}^1}+\beta\ket{\xi_{A''B''E}^2}$.
Taking into account the fact that $\ket{\psi_{ABE}'}=(U_A\otimes \mathbbm{1})\ket{\psi_{ABE}^2}$, we eventually arrive at (\ref{Lego}).

As a final remark, let us notice that the fact that the supports $V_1$ and $V_2$
are orthogonal implies also that $B_{1|i}$ are unitary because they are 
direct sums of $\widetilde{B}_i$ and $\widetilde{B}_i'$,
\begin{equation}\label{Block4}
    B_{1|i}=\left(
    \begin{array}{cc}
    \widetilde{B}_i & 0 \\
    0 & \widetilde{B}_i'
    \end{array}
    \right).
\end{equation}


\begin{thebibliography}{32}
\bibitem [{foo()}]{footnote1}%
  \BibitemOpen
  \bibinfo {note} {Let us note that $X_d^iZ_d^j$ are not always proper
  observables because their eigenvalues might differ from powers of $\omega$.
  However they can be easily turned into observables by including suitable
  multiplicative factors; in fact, one can show that
  $(X_d^iZ_d^j)^d=\omega^{ij(d-1)d/2}\mathbbm{1}$ for any $i,j$ and $d$. Note
  also that the statistics of $d^2$ observables $X_d^iZ_d^j$ can be obtained by
  measuring $d+1$ observables $X_dZ_d^j$ and $Z_d$ by changing the label of the
  outcomes.}\BibitemShut {Stop}%
\end{thebibliography}
\end{document}